\documentclass[preprint2]{aastex}

\newcommand{\Msun}{\hbox{$\hbox{M}_\odot\;$}}
\newcommand{\Rsun}{\hbox{$\hbox{R}_\odot\;$}}
\newcommand{\kms}{\hbox{${\rm km}\:{\rm s}^{-1}\;$}}
\newcommand{\Msuno}{\hbox{$\hbox{M}_\odot$}}
\newcommand{\Rsuno}{\hbox{$\hbox{R}_\odot$}}
\newcommand{\kmso}{\hbox{${\rm km}\:{\rm s}^{-1}$}}
\newcommand{\teff}{$T_{\rm eff}\;$}  
\newcommand{\teffo}{$T_{\rm eff}$}  
\newcommand{\logg}{$\log\;g\;$}  
\newcommand{\loggo}{$\log\;g$}

\slugcomment{To appear in ApJ}

\shorttitle{Stellar Abundances in V404 Cygni}
\shortauthors{J. I. Gonz\'alez Hern\'andez et al.}

\begin{document}


\title{Chemical Abundances of the Secondary Star \\ in the Black Hole
X-Ray Binary V404 Cygni}


\author{Jonay I. Gonz\'alez Hern\'andez\altaffilmark{1,2}, 
Jorge Casares\altaffilmark{1,2}, Rafael Rebolo\altaffilmark{1,2,3}, 
Garik Israelian\altaffilmark{1,2}, 
Alexei V. Filippenko\altaffilmark{4}, and 
Ryan Chornock\altaffilmark{4,5}}

\altaffiltext{1}{Instituto de Astrof{\'\i }sica de Canarias (IAC), 
E-38205 La Laguna, Tenerife, Spain; jonay@iac.es, jcv@iac.es, 
rrl@iac.es, gil@iac.es}
\altaffiltext{2}{Depto. Astrofísica, Universidad de La Laguna (ULL), 
E-38206 La Laguna, Tenerife, Spain} 
\altaffiltext{3}{Consejo Superior de Investigaciones 
Cient{\'\i}ficas, Spain} 
\altaffiltext{4}{Department of Astronomy, Uni\-ver\-si\-ty of
California, Berkeley, CA 94720-3411, USA; alex@astro.berkeley.edu} 
\altaffiltext{5}{Harvard-Smithsonian Center for Astrophysics, 
60 Garden Street, Cambridge, MA 02138, USA; rchornock@cfa.harvard.edu}


\begin{abstract}

We present a chemical abundance analysis of the secondary star in
the black hole binary \mbox{V404 Cygni}, using Keck~I/HIRES
spectra. We adopt a $\chi^2$-minimization procedure to
derive the stellar parameters, taking into account any possible veiling
from the accretion disk. With these parameters we determine the
atmospheric abundances of O, Na, Mg, Al, Si, Ca, Ti, Fe, and Ni. 
The abundances of Al, Si, and Ti appear to be slightly enhanced when
comparing with average values in thin-disk solar-type stars. The O 
abundance, derived from optical lines, is particularly enhanced in 
the atmosphere of the secondary star in \mbox{V404 Cygni}. 
This, together with the peculiar velocity of this system as 
compared with the Galactic velocity dispersion of thin-disk stars, 
suggests that the black hole formed in a supernova or hypernova 
explosion. We explore different supernova/hypernova models having
various geometries to study possible contamination of nucleosynthetic 
products in the chemical abundance pattern of the secondary star. 
We find reasonable agreement between the observed abundances and 
the model predictions. However, the O abundance seems to be too high 
regardless of the choice of explosion energy or mass cut, when 
trying to fit other element abundances. Moreover, Mg appears 
to be underabundant for all explosion models, which produces Mg 
abundances roughly 2 times higher than the observed value.

\end{abstract}


\keywords{black holes: physics --- stars: abundances --- stars: 
evolution --- stars: individual \mbox{V404 Cygni} ($=$\mbox{GS 2023+338}) --- 
supernovae: general --- X-rays: binaries}  

\section{Introduction}
       
The low-mass X-ray binary (LMXB) \mbox{V404 Cygni} is one of the
most compelling cases for a black hole accreting mass 
from a low-mass companion (secondary star). 
The mass function is among the highest values in such binary 
systems \citep[$f(M) = 6.08\pm 0.06$~\Msuno;][]{cas92,cas94},
placing the mass of the compact object comfortably above the 
upper limit of a rapidly rotating neutron star. The mass ratio of the
system, $q=M_2/M_{\rm BH}=0.060^{+0.004}_{-0.005}$, was derived from 
the measurement of the rotation velocity of the secondary star,
$v \sin i=39.1\pm 1.2$~\kms \citep{cas94}.   

\citet{sha94} modelled the ellipsoidal variations using a $K$-band infrared 
(IR) light curve of \mbox{V404 Cygni} and derived an orbital inclination
of $i = 56^\circ \pm 4^\circ$, and consequently the implied mass of the black
hole is $M_{\rm BH} = 12 \pm 2$~\Msuno. Later, \citet{pav96}
studied the $R$-band light curve to determine an inclination of 
$i = 56^\circ \pm 2^\circ$. More recently, \citet{kha10}
derived a K3~III spectral type for the secondary star from 
near-IR (NIR) spectroscopy of \mbox{V404 Cygni}, and
\citet{san96} remodeled the $H$-band light curve; they
refined the orbital inclination to be $i = (67^{+3}_{-1})^\circ$, which
yields a black hole mass of $M_{\rm BH} = 9^{+0.2}_{-0.6}$~\Msuno.

The system is placed near the Galactic plane, with a Galactic latitude 
of $b \approx -2.1^{\circ}$. \citet{mil09b} recently derived the 
distance with very long baseline interferometry (VLBI) observations, 
$d = 2.39 \pm 0.14$~kpc. They revisited the analysis of \citet{mil09a} 
to determine a more accurate peculiar velocity of $39.9 \pm 5.5$~\kmso. 
This value is significantly lower their previous one,
$v_{\rm pec}\sim 64$~\kms \citep{mil09a} assuming a distance 
of 4~kpc \citep{jon04}, and it can be achieved via a \citet{bla61} 
velocity kick; thus, no asymmetric supernova kick is 
required. However, \citet{mil09b} argued that the component of
the peculiar velocity in the Galactic plane, 39.6~\kmso, is still too
large to come from the intrinsic velocity dispersion of the Galactic
plane, which is 18.9~\kmso \citep{mig00} for the likely F0--F5 
progenitor of the donor star. Thus, this peculiar velocity
should come from a natal kick from the ejection of material in a
supernova (SN) or a more energetic hypernova (HN) event, but its 
magnitude does not require an additional asymmetric kick. 

The chemical abundances of secondary stars in black hole and 
neutron star X-ray binaries have been studied for several systems:
Nova Scorpii 1994 \citep{isr99,gon08a}, A0620--00 \citep{gon04}, 
Centaurus X-4 \citep{gon05b}, \mbox{XTE J1118+480}
\citep{gon06,gon08b}, and V4641 Sagittarii \citep{oro01,sad06}.
The metallicities of these binary systems are all close to
solar independent of their location with respect to 
the Galactic plane. In addition, the above authors have taken 
into account different scenarios of SN/HN ejecta pollution 
on the photospheric abundances of the secondary star. 

In this paper, we use high-resolution spectra to derive the 
stellar parameters and chemical abundances of the secondary 
star in the black hole X-ray binary \mbox{V404 Cygni}.
We then compare in detail these element abundances in the context of
the possible enrichment of the secondary star from SN/HN yields. 

\section{Observations}

We obtained 14 high-quality spectra of \mbox{V404 Cygni} in 
quiescence with the High Resolution
Spectrometer \citep[HIRES;][]{vog94} at the
Keck~I 10-m telescope (Mauna Kea, Hawaii) on 12 July 2009 UT. 
The data covered the spectral regions 4390--5805~{\AA}, 
5895--7405~{\AA}, and 7520--8780~{\AA} with a slit width of $0.861''$,
at a resolving power $\lambda/\Delta\lambda \approx 50,000$. The 
seeing was in the range $0.6''$--$1.0''$ during the whole night, 
and for most of the main target spectra it was $\sim 0.75''$.
This observing program was scheduled at a time such that the 
secondary star was near inferior conjunction with the black hole
(i.e., at an orbital phase of 0), to minimize possible effects 
on the determination of chemical abundances due to the asymmetry 
of its Roche lobe.

We also observed ten template 
stars with spectral types in the range K0V/IV--K2V with the same
instrument and spectral configuration. The integration time for
\mbox{V404 Cygni} was fixed at 1800~s in all exposures except for the
last two, which were of 2100~s and 2000~s duration. This relatively 
long integration time was chosen because the orbital smearing at 
this phase is only 3--5 \kmso, smaller than the instrumental 
resolution of $\sim 6.6$ \kmso. 

\begin{figure*}[ht!]
\centering
\includegraphics[width=13cm,angle=0]{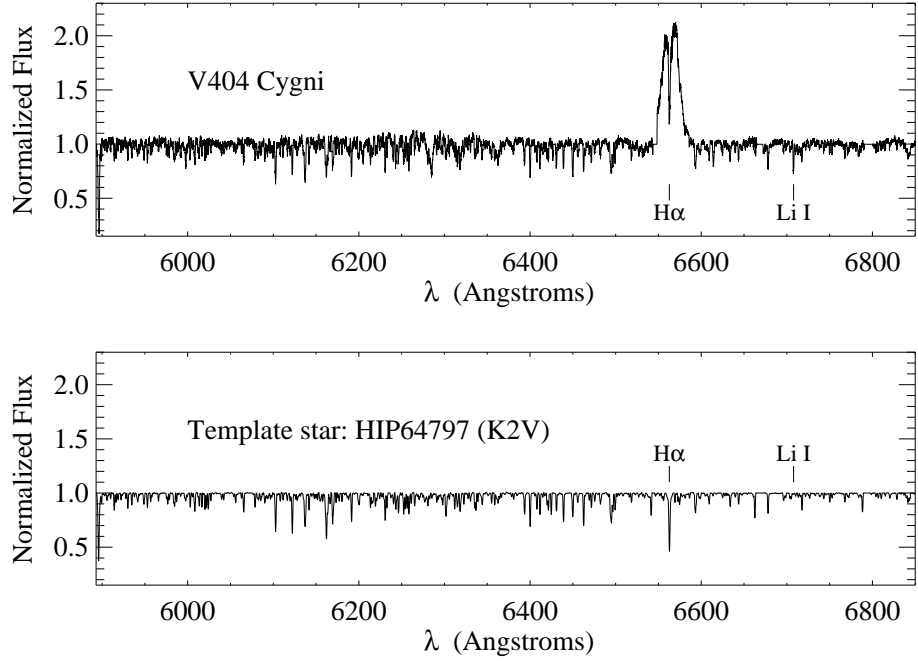}
\caption{\footnotesize{Observed spectrum of the secondary star of
\mbox{V404 Cygni}, after correcting for the individual radial
velocities and combining all individual spectra (top panel), 
and a properly broadened template also corrected for its radial
velocity (HIP~64797, bottom panel).}}  
\label{fspec}
\end{figure*}

The spectra were reduced in a standard manner using the 
{\scshape makee} package. Each order of
each individual spectrum was normalized using a low-order cubic spline 
and combined into a single one-dimensional spectrum.
The individual spectra were corrected for their radial velocity 
\citep[for more details, see][in preparation]{gon11} and combined 
in order to improve the signal-to-noise ratio (S/N). After
binning in wavelength in steps of 0.1~{\AA}, the final spectrum had 
S/N~$\approx 70$ in the continuum at 6500~{\AA} and S/N~$\approx 150$
at 7770~{\AA}. This spectrum is displayed in Figure~\ref{fspec},
in comparison with that of a K2~V template star. Note that the 
spectrum of the secondary star in this system and the spectra of 
the template stars were normalized using the same procedure.

\begin{deluxetable}{lccccc}
\tabletypesize{\scriptsize}
\tablecaption{Ranges and Steps of Model Parameters}   
\tablewidth{0pt}
\tablehead{\colhead{Parameter} & \colhead{Range} & \colhead{Step}}
\startdata
$T_{\mathrm{eff}}$  & $4200 \rightarrow 5800$ K & 100 K\\ 
$\log [g/({\rm cm~s}^2)]$  & $2 \rightarrow 4$  & 0.1\\ 
$\mathrm{[Fe/H]}$ & $-0.5 \rightarrow 0.6$  & 0.1\\ 
$f_{4500}$ &  $0 \rightarrow 0.6$  & 0.05\\ 
$m_0$  & $0 \rightarrow -0.000171$ & $-$0.000019\\ 
\enddata
\label{tpar}
\end{deluxetable}

\begin{figure*}[ht!]
\centering
\includegraphics[width=11cm,angle=0]{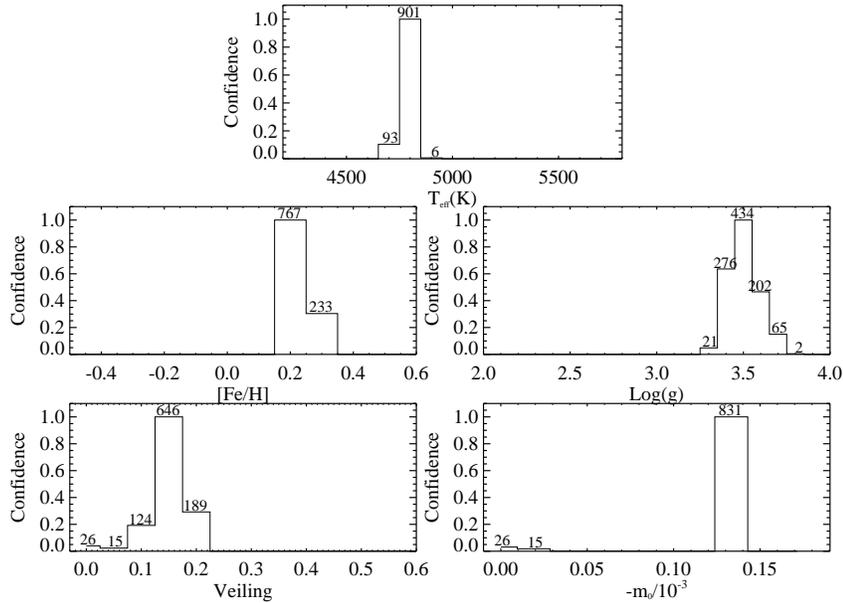}
\caption{\footnotesize{Distributions obtained for each parameter using
Monte Carlo simulations. The bottom-right panel shows the distribution
obtained for the veiling slope, $m_0$, given as $-m_0/10^{-3}$
 in units of \AA$^{-1}$.
The labels at the top of each bin indicate
the number of simulations consistent with the bin value. The total
number of simulations was 1000.}}
\label{fpar}
\end{figure*}

\begin{figure*}[ht!]
\centering
\includegraphics[width=9cm,angle=90]{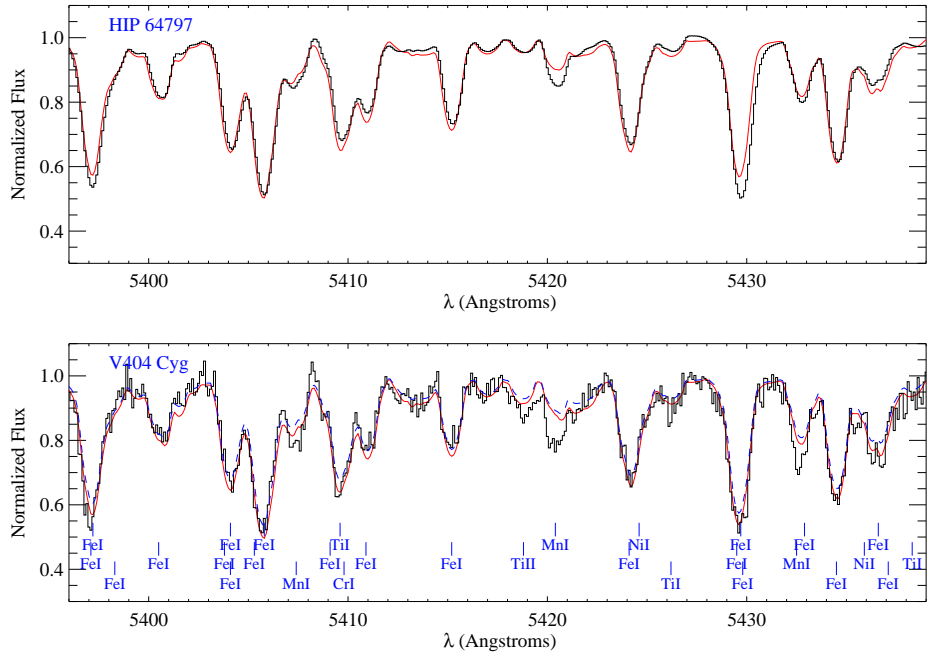}
\caption{\footnotesize{Best synthetic spectral fits to the 
Keck/HIRES spectrum of the secondary star in the \mbox{V404 Cygni} 
system (bottom panel), and the same for a template star 
(properly broadened) shown for comparison (top panel). 
Synthetic spectra are computed for solar abundances
(dashed line) and best-fit abundances (solid line).}}
\label{fsyn1}
\end{figure*}

\begin{figure*}[ht!]
\centering
\includegraphics[width=9cm,angle=90]{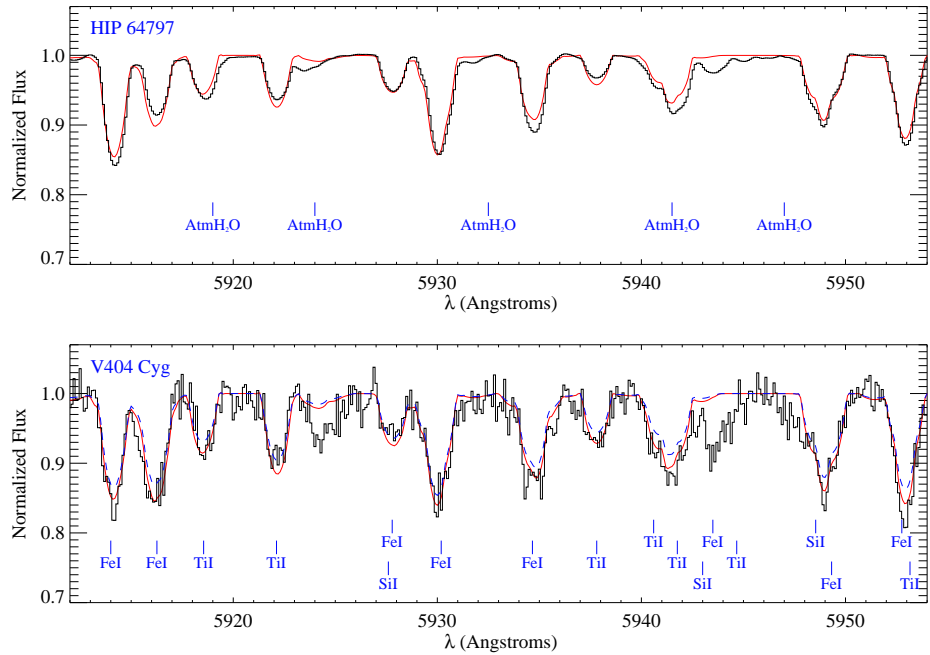}
\caption{\footnotesize{The same as in Fig.~\ref{fsyn1}, but for the 
spectral range 5910--5955~{\AA}. In this spectral region there 
are several relatively weak telluric lines labelled as AtmH$_2$O.}}  
\label{fsyn2}
\end{figure*}

\begin{figure*}[ht!]
\centering
\includegraphics[width=9cm,angle=90]{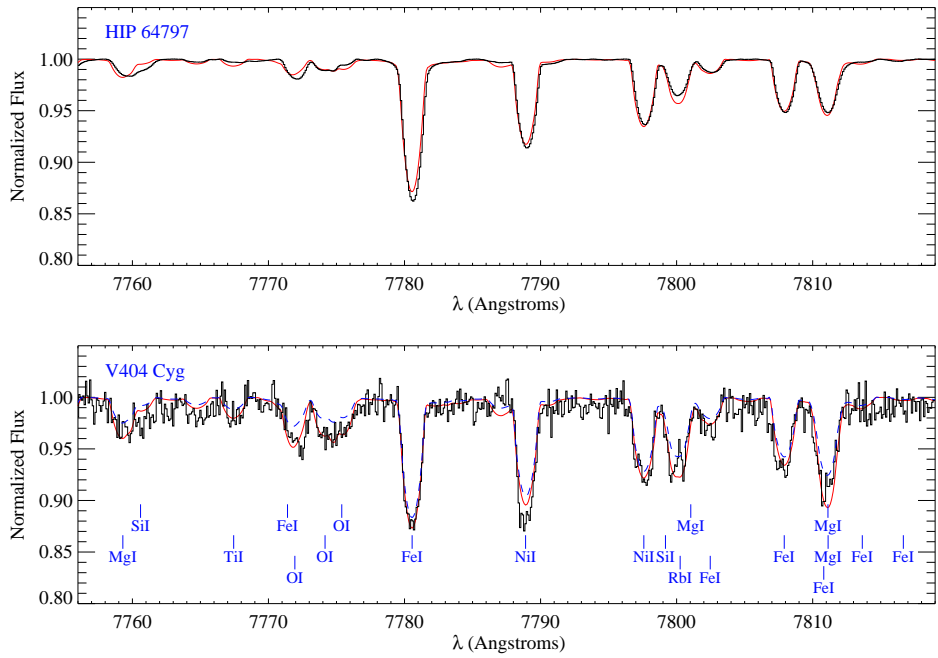}
\caption{\footnotesize{The same as in Fig.~\ref{fsyn1}, but for the 
spectral range 7756--7819~{\AA}.}}  
\label{fsyn3}
\end{figure*}

\section{Chemical Analysis}

\subsection{Stellar Parameters\label{spar}}

The merged and normalized spectrum of the secondary star in 
\mbox{V404 Cygni} may show apparently weaker stellar lines due to
the veiling introduced by the accretion disk. This veiling 
is found to be $\sim 13\% \pm 2\%$ in the range 6400--6600~{\AA} 
in a set of observations taken during the period 1990--1993 
\citep{cas94}. This veiling was estimated by performing standard 
optimal subtraction techniques with a K0~IV template star.

The high quality and resolution of the Keck/HIRES spectrum 
enable us to infer the stellar parameters, ($T_{\mathrm{eff}}$, 
$\log g$, and the metallicity [Fe/H]) of the companion star, 
taking into account any possible veiling from 
the accretion disk as in previous studies of other LMXBs
\citep[see, e.g.,][]{gon04,gon08b}. For simplicity, the veiling was 
defined as a linear function of wavelength and thus described with 
two additional parameters, the veiling at 4500~{\AA}, 
$f_{4500} = F^{4500}_{\rm disk}/F^{4500}_{\rm sec}$, and the slope,
$m_0$. We note that the total flux is defined as 
$F_{\rm total} = F_{\rm disk} + F_{\rm  sec}$, where 
$F_{\rm disk}$ and $F_{\rm sec}$ are the flux contributions of 
the disk and the continuum of the secondary star, respectively. 

The most recent version of this code \citep[see also][]{gon09} 
allows us to compare, via a $\chi^2$-minimization procedure,
up to 50 small spectral regions of the stellar spectrum with a grid 
of synthetic spectra computed using the local thermodynamical
equilibrium (LTE) code MOOG \citep{sne73}. 
We used a grid of LTE model atmospheres \citep{kur93} and the 
atomic line data from the Vienna Atomic Line Database
\citep[VALD][]{pis95}. The oscillator strengths of relevant
lines were adjusted until they reproduced the solar atlas
\citep{kur84} with solar abundances \citep{gre96}. The corrections
applied to the $\log gf$ values taken from the VALD database were
smaller than $\sim 0.2$ dex. 

\begin{deluxetable}{lrrrrrrrrrrrr}
\tabletypesize{\scriptsize}
\tablecaption{Chemical Abundances of the Secondary Star in V404 Cygni}
\tablewidth{0pt}
\tablehead{\colhead{Element} & $\log \epsilon(\mathrm{X})_{\odot}$\tablenotemark{a} & 
$[{\rm X}/{\rm H}]$ & $[{\rm X}/{\rm Fe}]$ & 
${\sigma}$ & $\Delta_\sigma$ &  
$\Delta_{T_{\rm eff}}$ & $\Delta_{\log g}$ &
$\Delta_\xi$ & $\Delta_{\rm vel}$ & $\Delta \mathrm{[X/H]}$ & $\Delta \mathrm{[X/Fe]}$ & 
$n$\tablenotemark{b}} 
\startdata
Fe &  7.50 &  0.23 &  --     &  0.12 &  0.02 &  0.05   &  0.00   & $-$0.16 &  0.08 &  0.19 &  --   & 51 \\ 
O\tablenotemark{d} &  8.74   &  0.60 &  0.37 &  0.07   &  0.05   & $-$0.16 &  0.06 & $-$0.06 &  0.06 &  0.20 &  0.24 &  2 \\ 
Na &  6.33 &  0.30 &  0.07   &  0.20 &  0.12 &  0.10   & $-$0.03 & $-$0.13 &  0.10 &  0.23 &  0.14 &  3 \\ 
Mg &  7.58 &  0.00 & $-$0.23 &  0.09 &  0.03 &  0.06   & $-$0.03 & $-$0.09 &  0.08 &  0.15 &  0.09 &  8 \\ 
Al &  6.47 &  0.38 &  0.15   &  0.07 &  0.05 &  0.05   &  0.00   & $-$0.08 &  0.05 &  0.11 &  0.11 &  2 \\ 
Si &  7.55 &  0.36 &  0.13   &  0.22 &  0.06 & $-$0.06 &  0.04   & $-$0.10 &  0.06 &  0.15 &  0.15 & 14 \\ 
Ca &  6.36 &  0.20 & $-$0.03 &  0.16 &  0.05 &  0.11   & $-$0.04 & $-$0.24 &  0.10 &  0.29 &  0.12 & 11 \\ 
Ti &  5.02 &  0.42 &  0.19   &  0.24 &  0.06 &  0.15   &  0.02   & $-$0.28 &  0.12 &  0.34 &  0.17 & 15 \\ 
Cr &  5.67 &  0.31 &  0.08   &  0.15 &  0.07 &  0.15   & $-$0.05 & $-$0.07 &  0.07 &  0.20 &  0.16 &  5 \\ 
Ni &  6.25 &  0.21 & $-$0.02 &  0.32 &  0.11 &  0.06   & $-$0.00 & $-$0.25 &  0.14 &  0.31 &  0.15 &  9 \\ 
\enddata

\tablecomments{Chemical abundances and uncertainties due to the 
uncertainties $\Delta_{T_{\rm eff}} = +100$\,K, 
$\Delta_{\log g} = +0.15$ dex,
$\Delta_\xi = +0.5~$\kmso, and $\Delta_{\rm vel} = 0.05$.}

\tablenotetext{a}{The solar element abundances were adopted from
\citet{gre96} for all elements except oxygen which was 
taken from \citet{ecu06}.} 

\tablenotetext{b}{The uncertainties from the dispersion of the best
fits to different features, $\Delta_\sigma$, are estimated using the
following formula: $\Delta_\sigma =\sigma/\sqrt{N}$, where $\sigma$
is the standard deviation of the measurements.}

\tablenotetext{c}{Number of spectral features of this element analyzed in
the star, or if there is only one, its wavelength.}

\tablenotetext{d}{The oxygen abundance has been corrected for NLTE
effects, $\Delta_{\rm NLTE} = -0.2$, using the NLTE computations in 
\citet{ecu06}.}

\label{tabun}
\end{deluxetable}

We inspected the high-quality HIRES spectrum of \mbox{V404 Cygni}, 
trying to search for \ion{Fe}{1}--\ion{Fe}{2} lines.
We finally selected 16 spectral features containing more than 50 Fe 
lines with excitation potentials between 1 and 5~eV. We emphasize 
that these Fe lines have a range of $\log gf$ values and are
spread over the whole spectral range 5100--6750~{\AA}, and we 
try to find the best-fit model for different veiling factors
and various veiling slopes. Therefore,
the strength of a given spectral feature depends not only on
the stellar parameters and metallicity of the star but also on the 
veiling factor appropriate for that feature at the corresponding 
wavelength.
In addition, due to the relatively high rotation velocity of the 
secondary star, these Fe lines were blended with stellar lines of 
other elements such as Ti and Ni. 
However, the main contributors to the selected features were always 
the Fe lines, and the abundances of other elements were in any case 
scaled with the metallicity of the model. 

The five free parameters were varied in the ranges given in 
Table~\ref{tpar}. The rotation velocity of the companion star 
was measured to be 36.4~\kms using the HIRES spectrum
\citep[for further details, see][in preparation]{gon11}, and a
limb darkening of $\epsilon = 0.65$ was adopted. 
The microturbulence, $\xi$, was computed using an experimental 
expression as a function of effective temperature and surface 
gravity \citep{all04}. The value for the best-fit model given below 
is $\xi=1.117$~\kmso.
 
We obtain as most likely values $T_{\mathrm{eff}} = 4800 \pm 100$~K, 
$\log [g/({\rm cm~s}^2)] = 3.50 \pm 0.15$, 
$\mathrm{[Fe/H]} = 0.20 \pm 0.17$, 
$f_{4500} = 0.15 \pm 0.05$, and $m_0 = -0.00013 \pm 0.00002$. 
The 1$\sigma$ uncertainties of the five free parameters were 
determined using 1000 realizations whose corresponding histograms 
are displayed in Figure~\ref{fpar}. 
Thus, we find a very small veiling, with 
$f_{5170} \approx 0.05$ for the spectral region of the 
\ion{Mg}{1}~b triplet at 5167--5183~{\AA} and almost zero for longer 
wavelengths. This veiling is lower than that measured during early
phases of quiescence \citep{cas93}. We emphasize 
that the linear function adopted to model the
behavior of the veiling is strictly valid only in the spectral range 
5000--6800~{\AA}, and therefore should not be extrapolated beyond
that wavelength range. On the other hand, we have analyzed several 
spectral ranges around 7750~{\AA}, 8450~{\AA}, and 8750~{\AA} by 
assuming a veiling $f_{\rm NIR} = 0$ for all of them, and the 
\ion{Fe}{1} lines in these regions seem to be well reproduced by 
the spectral synthesis. 
However, a veiling of 0 throughout the infrared
might not be correct, and one would have to analyze 
spectral lines in the infrared to check if this veiling, derived 
in the optical, holds for longer wavelengths.  

The stellar parameters of secondary stars in LMXBs are particularly
relevant for the determination of orbital inclinations, which are
typically derived from the modelling of ellipsoidal variations using
optical and NIR light curves at quiescence 
\citep[see, e.g.,][]{gel06}.  
\citet{sha94} derived $T_{\mathrm{eff}} \approx 4360$~K for the
companion star in their modelling of the $K$-band light curve of 
\mbox{V404 Cygni}. \citet{hyn09} suggest that its most likely spectral
is K0~III, which implies 
$T_{\mathrm{eff}} \approx 4570$~K using the temperature scale of giants in
\citet{bel99}. \citet{kha10}, however, propose a spectral type of
K3~III as the best template that matches their NIR broad-band 
spectra with little contribution from the accretion disk,
providing a veiling of $\sim 2\%$ and $\sim 3\%$ for their $H-$~and 
$K$-band spectroscopy. This spectral type lends to a 
$T_{\mathrm{eff}}\sim 4300$~K, according to their model. 
However, our spectroscopic value ($T_{\rm eff} \approx 4800$~K) may
require more contribution of the flux from the accretion disk in 
the NIR bands \citep[see, e.g.,][]{hyn05}, which may also imply a 
different orbital inclination, and therefore a different black hole
mass.

\begin{deluxetable}{lcccccccc}
\tabletypesize{\scriptsize}
\tablecaption{Surface Gravities in LMXBs}
\tablewidth{0pt}
\tablehead{\colhead{Name\tablenotemark{\star}} & $P_{\rm orb}$ &
$M_{{\rm CO},f}$ & $M_{2,f}$ & $q_f$ & $a_{c,f}$ & $R_L$ & 
$\log g({R_L})$ & $\log g({\rm spec})$ \\
& [days] & [\Msuno] & [\Msuno] &  & [\Rsuno] & [\Rsuno] & 
 & } 
\startdata
V404 Cygni  & $6.47129   \pm(7 \times 10^{-5})$ & $9.00\pm0.60$ & $0.54\pm0.08$ & $0.060\pm0.005$ & $30.98\pm0.75$ & $5.51\pm0.27$ & $2.69\pm0.03$ & $3.50\pm0.15$ \\
Cen X-4     & $0.6290522 \pm(4 \times 10^{-7})$ & $1.50\pm0.40$ & $0.30\pm0.09$ & $0.200\pm0.030$ & $ 3.76\pm0.38$ & $0.95\pm0.14$ & $3.96\pm0.02$ & $3.90\pm0.30$ \\
A0620--00   & $0.32301405\pm(1 \times 10^{-8})$ & $6.61\pm0.25$ & $0.40\pm0.05$ & $0.060\pm0.004$ & $ 3.79\pm0.05$ & $0.67\pm0.02$ & $4.38\pm0.02$ & $4.20\pm0.30$ \\
Nova Sco 94 & $2.62168   \pm(1.4\times10^{-4})$ & $6.59\pm0.45$ & $2.76\pm0.33$ & $0.419\pm0.028$ & $16.85\pm0.48$ & $5.17\pm0.24$ & $3.45\pm0.01$ & $3.70\pm0.20$ \\
XTE J1118   & $0.1699339 \pm(2\times10^{-7})$   & $8.30\pm0.28$ & $0.22\pm0.07$ & $0.027\pm0.009$ & $ 2.64\pm0.04$ & $0.37\pm0.05$ & $4.65\pm0.04$ & $4.60\pm0.30$ \\
\enddata

\tablenotetext{\star}{References: V404 Cygni: \citet{cas94}, \citet{cas96}, 
\citet{kha10}, and this work; 
Centaurus X-4: \citet{tor02}, \citet{gon05b}, \citet{cas07b},
  \citet{kha10}; 
A0620-00: \citet{gon04}, \citet{nei08}, \citet{gon10a},
   \citet{can10}; 
Nova Scorpii 1994: \citet{hoo98}, \citet{gon08a}, \citet{sha03}; 
XTE J1118+480: \citet{tor04}, \citet{gel06}, \citet{gon06},
  \citet{gon08b} and this work;}

\tablecomments{Comparison between LMXB surface gravity determined
spectroscopically, $\log g({\rm spec})$, with
the surface gravity, $\log g({R_L})$, computed from the orbital
period, $P_{\rm orb}$, the mass ratio, $q_f$, and the current masses
of the compact object, $M_{{\rm CO},f}$, and the secondary star, 
$M_{2,f}$.}

\label{tgrav}
\end{deluxetable}

\begin{deluxetable}{lccccccc}
\tabletypesize{\scriptsize}
\tablecaption{Kinematical properties in LMXBs}
\tablewidth{0pt}
\tablehead{\colhead{Name\tablenotemark{\star}} & $K_2$ & $f(M)$ & 
$v \sin i$ & $\gamma$ & d & $\mu_\alpha$ & $\mu_\delta$ \\
 & [\kmso] & [\Msuno] & [\kmso] & [\kmso] & [Kpc] & [mas yr$^{-1}$] & [mas yr$^{-1}$] } 
\startdata
V404 Cygni  & $208.4\pm0.6$ & $6.09\pm0.04$   & $40.8\pm0.9$	& $0.3\pm0.6$	& $2.39\pm 0.14$ & $-5.04\pm0.02$ & $-7.64\pm0.03$ \\
Cen X-4     & $144.6\pm0.3$ & $0.197\pm0.001$ & $44\pm3$	& $189.6\pm0.2$ & $1.30\pm 0.40$ & $11\pm10$	  & $-56\pm10$  \\
A0620--00   & $435.4\pm0.5$ & $3.10\pm0.04$   & $82\pm2$	& $8.5\pm1.8$	& $1.06\pm 0.12$ & --	   & -- \\
Nova Sco 94 & $226.1\pm0.8$ & $3.14\pm0.03$   & $86\pm4$	& $-141.9\pm1.3$& $3.20\pm 0.50$ & $-3.3\pm 0.5$ & $-4.0\pm0.4$ \\
XTE J1118   & $708.8\pm1.4$ & $6.27\pm0.04$   & $100^{+3}_{-11}$& $2.7\pm1.1$	& $1.70\pm 0.10$ & $-16.8\pm1.6$ & $-7.4\pm1.6$ \\
\enddata

\tablenotetext{\star}{References: V404 Cygni: \citet{cas96}, \citet{cas07a}, \citet{mil09b}, \citet{kha10}; 
Centaurus X-4: \citet{gon05a}, \citet{cas07b}, \citet{kha10}; 
A0620-00: \citet{nei08}, \citet{gon10a}, \citet{can10}; 
Nova Scorpii 1994: \citet{hje95}, \citet{oro97}, \citet{mir02}, \citet{gon08a}, and this work; 
XTE J1118+480: \citet{mir01}, \citet{gel06}, \citet{gon08b}.}

\tablecomments{Kinematical and dynamical properties, orbital parameters 
and proper motions of low-mass X-ray binaries.}

\label{tgrav}
\end{deluxetable}

\begin{deluxetable}{lccccc}
\tabletypesize{\scriptsize}
\tablecaption{Stellar, veiling parameters and chemical abundances in LMXBs}
\tablewidth{0pt}
\tablehead{\colhead{Name\tablenotemark{\star}} & A0620--00 & Centaurus X-4 & XTE J1118+480 & Nova Sco 94 & 
V404 Cygni\tablenotemark{\dagger}} 
\startdata
Alternative name        & V616 Mon      &  V822 Cen     &  --           & GRO J1655--40  & GS 2023+338   \\ 
$T_{\mathrm{eff}}$ (K)  & $4900\pm100$  & $4500\pm100$  & $4700\pm100$  & $6100\pm200$   & $4800\pm100$  \\ 
$\log (g/{\rm cm~s}^2)$ & $4.2\pm0.3$   & $3.9\pm0.3$   & $4.6\pm0.3$   & $3.7\pm0.2$    & $3.50\pm0.15$ \\
$f_{4500}$              & $0.25\pm0.05$ & $1.85\pm0.10$ & $0.85\pm0.20$ & $0.15\pm0.05$  & $0.15\pm0.05$ \\
$m_0/10^{-4}$           & $-1.4\pm0.2$  & $-7.1\pm0.3$  & $-2\pm1$      & $-1.2\pm0.3$   & $-1.3\pm0.2$  \\
{[O/H]}\tablenotemark{\ddagger} & --   & --		& --		& $0.91\pm0.09$  & $0.60\pm0.19$ \\
{[Na/H]}	        & --            & --		& --		& $0.31\pm0.26$  & $0.30\pm0.19$ \\
{[Mg/H]}	        & $0.40\pm0.16$ & $0.35\pm0.17$ & $0.35\pm0.25$ & $0.48\pm0.15$  & $0.00\pm0.11$ \\
{[Al/H]}	        & $0.40\pm0.12$ & $0.30\pm0.17$ & $0.60\pm0.20$ & $0.05\pm0.18$  & $0.38\pm0.09$ \\
{[Si/H]}	        & --            & --		& $0.37\pm0.21$ & $0.58\pm0.08$  & $0.36\pm0.11$ \\
{[S/H]}                 & --            & --		& --            & $0.66\pm0.12$  &  --     \\
{[Ca/H]}	        & $0.10\pm0.20$ & $0.21\pm0.17$ & $0.15\pm0.23$ & $-0.02\pm0.14$ & $0.20\pm0.16$\\
{[Ti/H]}	        & $0.37\pm0.23$ & $0.40\pm0.17$ & $0.32\pm0.26$ & $0.27\pm0.22$  & $0.42\pm0.20$ \\
{[Cr/H]}	        & --            & --            & --            & --             & $0.31\pm0.19$ \\
{[Fe/H]}                & $0.14\pm0.20$ & $0.23\pm0.10$ & $0.18\pm0.17$ & $-0.11\pm0.10$ & $0.23\pm0.09$\\
{[Ni/H]}	        & $0.27\pm0.10$ & $0.35\pm0.10$ & $0.30\pm0.21$ & $0.00\pm0.21$	 & $0.21\pm0.19$	 \\
\enddata

\tablenotetext{\star}{References: V404 Cygni: This work; 
Centaurus X-4: \citet{gon05b}, \citet{gon07},; 
A0620-00: \citet{gon04}, \citet{gon07}; 
Nova Scorpii 1994: \citet{gon08a}; 
XTE J1118+480: \citet{gon08b}}

\tablenotetext{\dagger}{The uncertainties on the stellar abundances 
given in this table has been derived without taking into the error on 
the microturbulence (see Table~\ref{tabun}).}

\tablenotetext{\ddagger}{Oxygen abundances are given in NLTE.}

\tablecomments{Stellar and veiling parameters, and chemical 
abundances of secondary stars in low-mass X-ray binaries.}

\label{tgrav}
\end{deluxetable}

We have collected the most updated information on orbital parameters 
for several X-ray binaries from the literature to derive the 
expected size of the Roche lobe
and thus the corresponding surface gravity. We compare these values 
with our spectroscopic determinations in Table~\ref{tgrav}. 
Using Kepler's third law, we have estimated the current orbital 
separation, $a_{c,f}$, from the orbital period, $P_{\rm orb}$, 
and the reported masses of the compact object, $M_{{\rm CO},f}$, 
and the companion star, $M_{2,f}$.
The ratio of the radius of the Roche lobe, $R_L$, and the orbital 
separation can be estimated using Eggleton's expression \citep{egg83} 
as $$R_L/a_{c,f}=0.49q^{2/3}/[0.6q^{2/3}+\ln{(1+q^{1/3})}].$$ 
Assuming that the star is filling its Roche lobe, $R_L=R_{2,f}$, and 
using the mass of the secondary star, we derive the expected 
surface gravity.

From Table~\ref{tgrav} one can see that in general, the spectroscopic 
determinations are close to values derived from the size of 
the Roche lobe of the secondary star. In particular, for 
\mbox{Centaurus X-4}, \mbox{A0620-00}, and \mbox{XTE J1118+480},
the values are consistent within the 1$\sigma$ uncertainties, and
for \mbox{Nova Sco 94} at 1.2$\sigma$. However, for 
\mbox{V404 Cygni},
the surface gravity is only compatible at the 5$\sigma$ level, 
given the relatively small uncertainty in the spectroscopic 
surface gravity.
On the other hand, we note that surface gravities derived from the 
size of the Roche lobe strongly depend on the current masses of 
the compact object and the secondary star, and thus depend on the 
estimated, sometimes uncertain, orbital inclination.
We note, for instance, the
case of \mbox{A0620-00}, whose black hole mass estimate has changed 
over the last ten years from $11.0\pm1.9$ \Msun \citep{gel01}, to 
$9.7\pm0.6$ \Msun \citep{fro07}, and finally to $6.61 \pm 0.25$ 
\Msun \citep{can10}.
However, for the case of \mbox{V404 Cygni}, the mass ratio and
the orbital period have been determined with high precision (see
Table~\ref{tgrav}), leaving little room to increase the value 
of the surface gravity obtained from the secondary's mass and the 
size of the Roche lobe.
Thus, adopting larger masses \citep{sha94} for the compact object, 
$M_{{\rm CO},f}=12$~\Msuno, and the companion star, 
$M_{2,f}=0.7$~\Msuno, the result does not change 
significantly; the estimated surface gravity would be 
$\log g_{R_L} = 2.72\pm0.13$. Therefore, a possible uncertain orbital
inclination does not seem to be the reason for this disagreement. 
 
\subsection{Stellar Abundances\label{secabun}}

We inspected several regions in the observed Keck/HIRES 
spectrum of the secondary star, searching for suitable lines for 
a detailed chemical analysis. Using the derived stellar parameters, 
we first determined the Fe abundance by comparing synthetic 
spectra with each individual feature in the HIRES spectrum 
(see Table~\ref{tabun}). In Figure~\ref{fsyn1} we display one of the 
spectral regions analyzed to obtain the Fe abundance, also showing 
the best synthetic spectral fit to the observed spectrum of a 
template star (HIP~64797 with \teffo $=4970$~K, 
\loggo $= 4.33$, and [Fe/H] $=-0.24$ dex). We only use as abundance
indicators those features which are well reproduced in the template
star. The chemical analysis is summarized in Table~\ref{tabun}. The
errors in the element abundances show their sensitivity to the
uncertainties in the effective temperature
($\Delta_{T_{\mathrm{eff}}}$), gravity ($\Delta_{\log g}$), veiling
($\Delta_{\rm vel}$), microturbulence ($\Delta_\xi$), and 
the dispersion of the measurements from different spectral 
features ($\Delta_\sigma$). In
Table~\ref{tabun} we also state the number of features analyzed for
each element.  
The uncertainties $\Delta_\sigma$ were estimated as $\Delta_\sigma
=\sigma/\sqrt{N}$, where $\sigma$ is the standard deviation of the $N$
measurements. The uncertainties $\Delta_{T_{\mathrm{eff}}}$, $\Delta_\xi$,
$\Delta_{\log g}$, and $\Delta_{\rm vel}$ were determined in the same
way as, for instance, in the \teff case: $\Delta_{T_{\rm
eff}} = (\sum_{i=1}^N\;\Delta_{T_{\rm eff},i})/N$. 
The total uncertainty given in Table~\ref{tabun} was derived using the 
following expression:
$\Delta \mathrm{[X/H]} = \sqrt{\Delta_{\sigma}^2 +
\Delta_{T_{\mathrm{eff}}}^2 + \Delta_{\log g}^2 + \Delta_{\xi}^2 +
 \Delta_{\rm vel}^2}$.

In Figure~\ref{fsyn2} we show the spectral region 5910--5955~{\AA},
where there are some Ti and Si lines used to derive their
abundances. In this region there are also some telluric lines, with 
equivalent widths of $\sim 30$~m{\AA}, which are significantly 
weaker than the stellar features, whose equivalent widths are
$\sim 250$~m{\AA}, and we do not think these lines have caused 
any problem in the abundance determination. This spectral region
contains two Fe features at $\sim 5914.2$ and $\sim 5916.3$~{\AA} 
which are sensitive to surface gravity variations. 
One can easily notice the relative different strengths of
both Fe features in the template spectra (with \loggo $\approx 4.3$) 
and in the secondary star in \mbox{V404 Cygni} (with 
\loggo$ \approx 3.5$).

\subsection{Oxygen}

The oxygen abundance was derived from the \ion{O}{1} triplet at 
7771--5~{\AA}, which for the relatively high rotation velocity of 
the secondary star in \mbox{V404 Cygni} produces two well-resolved
features that we have analyzed independently. In Figure~\ref{fsyn3} we
show the 7755--7820~{\AA} range, where the \ion{O}{1} features in the
secondary star appear to be apparently enhanced when compared with
those in the template star having similar effective
temperature. The best fit in LTE gives an oxygen abundance of 
$\mathrm{[O/H]_{\rm LTE}} \approx 0.8$. We note that the atomic data 
for \ion{O}{1} lines were adopted from \citet{ecu06}. These authors 
slightly modified the oscillator strengths in order to obtain a solar 
oxygen abundance of $\log \epsilon(\mathrm{O})_{\odot}=8.74$. For
other elements we assumed as solar abundances those given by
\citet{gre96}.

The \ion{O}{1} $\lambda$7771--5 triplet suffers from appreciable 
non-LTE (NLTE) effects \citep[see, e.g.,][]{ecu06}. For the stellar
parameters and oxygen abundance of the secondary star, NLTE 
corrections\footnote{$\Delta_\mathrm{NLTE} = \log 
\epsilon(\mathrm{X})_\mathrm{NLTE}-\log 
\epsilon(\mathrm{X})_\mathrm{LTE}$.} are estimated to be $\sim
-0.20$~dex for the NIR \ion{O}{1} $\lambda7771$--5 triplet. 
Table~\ref{tabun} provides the oxygen abundance properly
corrected for NLTE effects. 

\begin{figure*}[ht!]
\centering
\includegraphics[width=13.5cm,angle=0]{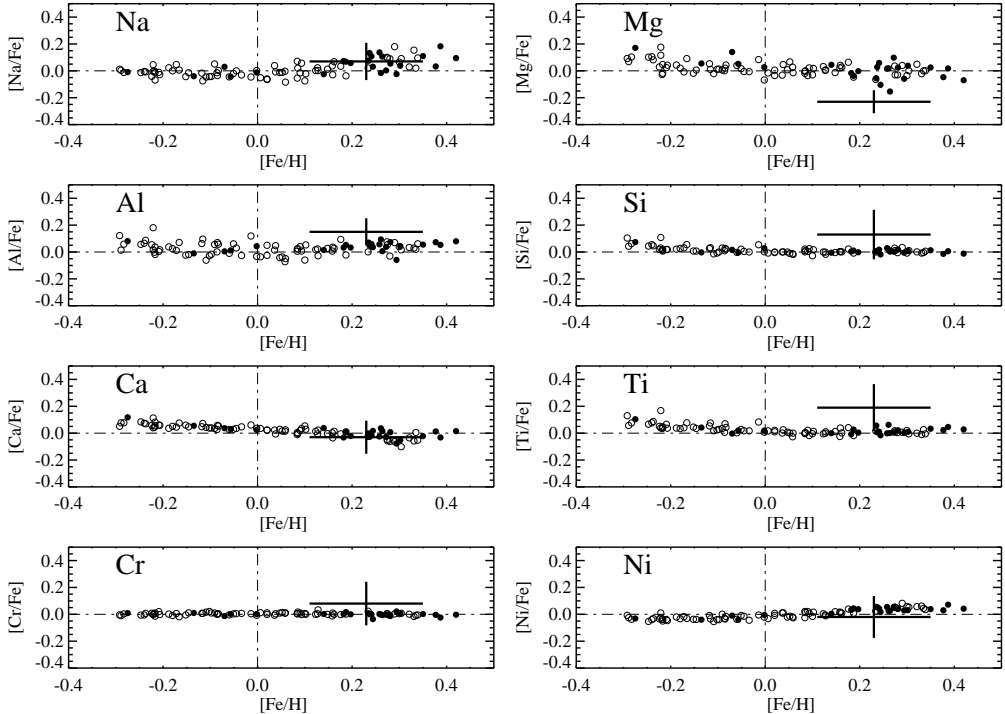}
\caption{\footnotesize{Abundance ratios of the secondary star in
\mbox{V404 Cygni} (blue wide cross) in comparison with the
abundances of solar-type metal-rich dwarf stars. Galactic trends were
taken from \citet{gon10b}. The size of the cross 
indicates the uncertainty. Filled and empty circles correspond to
abundances for exoplanet host stars and stars without known exoplanet
companions, respectively. The dashed-dotted lines indicate solar
abundance values.}} 
\label{fabfe}
\end{figure*}

In principle, one could argue that the relatively ``high'' value 
for the spectroscopic surface gravity found in this work (see
\S \ref{spar}) for \mbox{V404 Cygni} may affect the derived
abundances. We note that the
spectroscopic values were derived for the secondary star at
nearly inferior conjunction, which means that we are looking at the
``spherical'' side of the Roche-lobe-like star.
Therefore, the derived \teff and \logg 
values should be larger than the mean \teff and \logg values of 
the star (see Table~\ref{tgrav}), although this would only account 
for 100~K and 0.1~dex, respectively. 

In Table~\ref{tabun} one can see that the NIR \ion{O}{1} lines are sensitive
to the surface gravity, with a change of 0.06~dex in derived abundance for 
a change of $+0.15$~dex in $\log g$. This means that for a surface
gravity as low as $\log g = 2.7$, one would obtain a 0.32~dex lower
oxygen abundance. On the other hand, the Fe abundance would not be
sensitive to this change in surface gravity, and therefore we would
get $[{\rm O}/{\rm Fe}]_{\rm NLTE}= 0.05$~dex.\footnote{Fe lines 
are indeed quite sensitive to variations of the surface gravity, 
typically with changes of $\pm 0.05$~dex in Fe abundance for a 
change of $+0.15$~dex in $\log g$, although some Fe lines would 
not show any change in abundance. However, these differences 
compensate each other, so that the average change is 
$\sim 0.001$~dex, almost null.} We also note that, 
for instance, a 300~K lower \teff would result in a substantial 
increase in the derived oxygen abundance, $\sim 0.48$~dex (see 
Table~\ref{tabun}), and a smaller decrease in the Fe abundance, 
$\sim -0.15$~dex, yielding $[{\rm O}/{\rm Fe}]_{\rm NLTE}= 1$~dex
with $[{\rm Fe}/{\rm H}]= 0.08$. 
These values has been estimated without taking 
into account the possibly smaller NLTE correction at the lower
value of \teffo.

\subsection{Magnesium}
  
The magnesium abundance was derived from eight \ion{Mg}{1}
features, including the four optical lines of the 
\ion{Mg}{1}~b~$\lambda$5167--83~{\AA}~triplet and
\ion{Mg}{1}~$\lambda$5528~{\AA}. These \ion{Mg}{1} lines are known to be 
sensitive to NLTE effects \citep[see, e.g.,][]{zha98},  
although with NLTE corrections of only $\sim+0.05$ in the Sun and 
probably smaller ones at these cooler 
temperatures \citep{zha00}. 

We note here that we are applying an automatic 1.5$\sigma$ 
abundance rejection over the initial set of lines, 
slightly affecting the Mg abundance. This discards two \ion{Mg}{1} 
features, lowering the Mg abundance by 0.06~dex and decreasing the
standard deviation of the measurements from 0.15~to 
0.09~dex. For other elements like Si and Ti, the 1.5$\sigma$
rejection yields 0.04 lower and 0.03 higher abundance, respectively,
with a decrease of the standard deviation by only 0.02~dex.

Finally, possible differences in the stellar parameters also have
an impact on the Mg abundance (see Table~\ref{tabun}).
The \ion{Mg}{1} lines are sensitive to the effective temperature
and surface gravity, with a change of 0.06 and $-$0.03~dex in derived 
abundance for changes of 100~K in \teff and $+0.15$~dex in \loggo,
respectively. In this case, 300~K lower \teff would result in 0.18~dex
lower Mg abundance, whereas a decrease of 0.8~dex in \logg would 
give rise to a 0.16~dex higher Mg abundance. 

\section{Discussion}

\begin{deluxetable}{lccccc}
\tabletypesize{\scriptsize}
\tablecaption{Element Abundance Ratios in \mbox{V404 Cygni}}
\tablewidth{0pt}
\tablehead{\colhead{Element} & $\mathrm{[X/Fe]}_{\rm V404Cyg}$ &
$\Delta^{\tablenotemark{\star}}_{\rm [X/Fe],V404Cyg}$ & $\mathrm{[X/Fe]}_{\rm stars}$ & $\sigma_{\rm
stars}$ & $\Delta_{\sigma,{\rm stars}}$} 
\startdata
O    &  0.37   & 0.24 & $-$0.32 & 0.19 & 0.033 \\
Na   &  0.07   & 0.14 &  0.05	& 0.05 & 0.009 \\
Mg   & $-$0.23 & 0.09 &  0.00	& 0.05 & 0.008 \\
Al   &  0.15   & 0.11 &  0.04	& 0.03 & 0.006 \\
Si   &  0.13   & 0.15 &  0.00	& 0.01 & 0.002 \\
Ca   & $-$0.03 & 0.12 & $-$0.02 & 0.03 & 0.005 \\
Ti   &  0.19   & 0.17 &  0.01	& 0.02 & 0.003 \\
Cr   &  0.08   & 0.16 &  0.00	& 0.01 & 0.001 \\
Ni   & $-$0.02 & 0.15 &  0.03	& 0.02 & 0.003 \\
\enddata

\tablenotetext{\star}{Uncertainties in the element abundance ratios
($\mathrm{[X/Fe]}$) in the secondary star in \mbox{V404 Cygni}.}

\tablecomments{$\mathrm{[X/Fe]}_{\rm stars}$ indicates the average
value of 36 stars with iron content in the range 0.11 to 0.35
corresponding to the abundance range 
$\mathrm{[Fe/H]}\pm\sigma_\mathrm{Fe}$
of the secondary star in \mbox{V404 Cygni}, taken from 
\citet{ecu06} and \citet{gon10b}. 
The uncertainty in the average value of an abundance ratio in the 
comparison sample is obtained as $\Delta_{\sigma,{\rm
stars}} =\sigma_{\rm stars}/\sqrt{N}$, where $\sigma_{\rm stars}$ is
the standard deviation of the measurements and $N$ is the number of 
stars.}    

\label{tabfe}
\end{deluxetable}

\begin{figure}[ht!]
\centering
\includegraphics[width=7.8cm,angle=0]{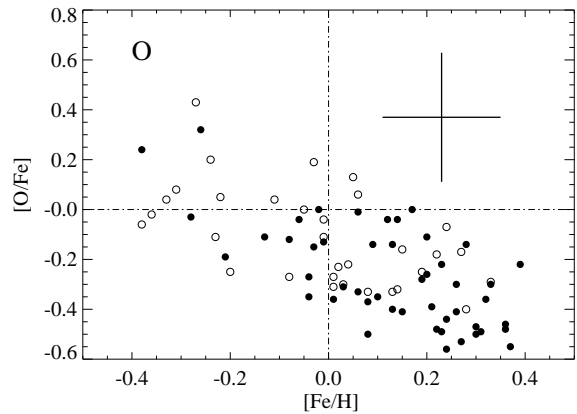}
\caption{\footnotesize{The same as in Fig.~\ref{fabfe} but for oxygen.
Galactic trends were taken from \citet{ecu06}.}} 
\label{fabO}
\end{figure}

The global metallicity of the secondary star in \mbox{V404 Cygni}
is slightly higher than solar, similar to that of the secondary
star in other black hole X-ray binaries such as \mbox{A0620--00} and 
\mbox{XTE J1118+480}, but 
also comparable to that of many stars in the solar neighborhood. 
We have searched for anomalies in
the abundance pattern of the secondary star in comparison with typical
abundances of stars belonging to the Galactic thin disk.
The abundances of other elements relative to iron listed in
Table~\ref{tabun} are compared in Figures~\ref{fabfe} and~\ref{fabO} with
the Galactic trends of these elements in the relevant range of
metallicities. We adopted recent, very accurate Galactic trends 
from \citet{gon10b} for all elements except for oxygen, because we 
needed O abundances measured from the NIR \ion{O}{1} triplet. We thus
decided to adopt the Galactic trend of oxygen published by \citet{ecu06}.
Figures~\ref{fabfe} and~\ref{fabO} show moderate anomalies for Ti, Al, 
and Si in the secondary star in \mbox{V404 Cygni}, and whereas Mg is 
surprisingly underabundant, oxygen appears to be enhanced 
in comparison with the galactic trend. 

Table~\ref{tabfe} shows the element abundance ratios 
in \mbox{V404 Cygni} and the average 
values in stars with iron content in the range $0.11 < 
\mathrm{[Fe/H]} < 0.35$, the
comparison sample corresponding to a 1$\sigma$ uncertainty in the
iron abundance of the secondary star. The elements Na, Si, Ca, Cr,
and Ni are consistent with the average values in thin-disk stars 
within the error bars. Mg appears to be underabundant, whereas Al and Ti
are roughly consistent at 1$\sigma$. Oxygen may be considered
separately since it is clearly more abundant, at 2.8$\sigma$, 
than the average values of the thin-disk stars.

\mbox{V404 Cygni} is located in a Galactic thin-disk region at 
a distance $d = 2.39 \pm 0.14$~kpc from
the Sun \citep{mil09b}, but its peculiar velocity is larger than
the intrinsic velocity dispersion in the Galactic plane. \citet{mil09b}
argue that this peculiar velocity, $v_{\rm pec}=39.9 \pm 5.5$~\kmso, 
may be the result of the natal kick from the mass ejection in the 
SN/HN explosion of the primary star in this LMXB.
The companion star could have captured a significant amount of 
the ejected matter in the SN/HN explosion that formed the 
compact object. We have explored this possibility with the aim of
obtaining information on the chemical composition of the progenitor 
of the compact object.

\subsection{Spherical Explosion}

The secondary star in \mbox{V404 Cygni} may have lost a 
significant amount of its initial mass through a mass-transfer 
mechanism onto the compact object, during the binary evolution. 
Its post-SN evolution has been studied in detail
by \citet{mil09a}. These authors suggest that from
0.5 to 1.5 \Msun may have been accreted from the companion star. 
We estimated the current mass of the secondary star at  
$M_{2,f}=0.54 \pm 0.08$~\Msun (see Table~\ref{tgrav}), from the 
current black hole mass $M_{{\rm BH},f}=9.0 \pm 0.6$~\Msun 
\citep{kha10} and mass ratio $q_f=0.060 \pm 0.005$ \citep{cas94}. 
Following \citet{mil09a}, we assume an initial mass for the 
secondary star of $M_{2,i} \approx 2$~\Msun and 
an initial black hole mass of $M_{{\rm BH},i} \approx 8$~\Msuno. 
Therefore, the final mass of the black hole would be 
$M_{{\rm BH},f} \approx 9.5$~\Msun which is consistent with
the current estimate of the black hole mass \citep{kha10}.

The high O content in this companion star may not suggest
any strong CNO processing within the star itself during
its evolution, since this would increase the N abundance
\citep[see, e.g.,][]{has02}. Oxygen, however, is not expected 
to decrease too much compared withC \citep{cla83}. 
Therefore, the CNO-processed material is C underabundant 
and N overabundant. Unfortunately, in the Keck/HIRES optical spectrum 
of the secondary star there are no sufficiently strong and unblended 
stellar lines to provide an accurate and reliable C abundance. 
\citet{kha10} argued that their NIR spectrum of \mbox{V404 Cygni} 
suggests that CO molecules seems to fit well with a solar-abundance 
K3~III template spectrum for which they adopt $T_{\rm eff} \approx 
4300$~K. We note here that CO molecular bands are more sensitive to 
the C abundance than to the O abundance and are stronger at lower 
effective temperatures. Thus, their argument do not necessarily 
contradict our determination of the O abundance.

The present orbital distance is $a_{c,f} \approx 31$~\Rsun 
(see Table~\ref{tgrav}).  
We will assume that the post-SN orbital separation after tidal 
circularization of the orbit was $a_{c,i} \approx 15$~\Rsuno, 
since the secondary must have experienced
mass and angular momentum losses during the binary evolution until
it reached its present configuration.
Similarly to the case of \mbox{XTE J1118+480} \citep{gon08b}, 
we can estimate the maximum ejected mass in the SN/HN explosion.
A binary system such as \mbox{V404 Cygni} will
survive a spherical SN explosion if the ejected mass $\Delta
M = M_{\rm He}-M_{{\rm BH},i} \le (M_{\rm He}+M_{2,i})/2$ \citep{hil83}.
This implies a mass of the He core before the SN explosion of 
$M_{\rm He} \le 18$ \Msuno. We therefore adopt He core masses of 
$M_{\rm He} \approx 15$--16~\Msuno, so we infer a He core radius
of $R_{\rm He} \approx 3$~\Rsun from the expression in 
\citet[][and references therein]{por97a}.  

Assuming a pre-SN circular orbit and an instantaneous spherically
symmetric ejection (that is, shorter than the orbital period), one can
estimate the pre-SN orbital separation, $a_0$, using the relation
given by \citet{heu84}: $a_0=a_{c,i}\,\mu_f$, where
$\mu_f=(M_{{\rm BH},i}+M_{2,i})/(M_{\rm He}+M_{2,i})$. We find $a_0
\approx 9$, for the adopted values of $M_{\rm He}=15$~\Msun 
and $a_{c,i}=15$~\Rsuno. 
At the time of the SN explosion \citep[$\sim$5--6 Myr;][]{bru82}, 
a 2~\Msun secondary star, still in its pre-main-sequence 
evolution, has a radius $R_{2,i} \approx 3.3$~\Rsuno. 
The material possibly captured by the secondary star in the SN/HN event
has a much larger mean molecular weight than the pre-explosion gas of the
secondary star, so it has probably been well mixed within the star
by thermohaline mixing in a relatively short timescale 
\citep[][and references therein]{pod02}. Therefore, the subsequent lost
of material from the secondary star onto the compact object should not
affect the chemical composition of its atmosphere.
The amount of mass deposited on the secondary can be estimated as 
$m_{\rm cap}=\Delta M (\pi R_{2,i}^2/4 \pi a_0^2)f_{\rm cap}$~\Msuno, 
where $f_{\rm cap}$ is the fraction of mass, ejected within the solid 
angle subtended by the secondary star, that is eventually captured. We
assume that the captured mass, $m_{\rm cap}$, is completely mixed with
the rest of the companion star. 

We compute the expected abundances in the atmosphere of the secondary
star after the pollution from the progenitor of the compact
object as in \citet{gon04} and \citet{gon08b}. We use 
40~\Msun spherically symmetric core-collapse explosion models 
($M_{\rm He} \approx 15.1$~\Msuno) at solar metallicity ($Z=0.02$)
and for two different explosion energies \citep{ume02,ume05,tom07}. 
These models imply $\Delta M \approx 7$~\Msun and need small capture 
efficiencies of $f_{\rm cap} \la 0.1$ (i.e., 10\%) to increase 
significantly the metal content of the secondary star. In order to 
fix the $f_{\rm cap}$ parameter, we have tried to get an expected Al
abundance, $[{\rm Al/H}] \approx 0.45$, consistent with the 
observed Al abundance within the error bars. 
These model computations are shown in Table~\ref{tsn}. 
These models would also provide a different mass fraction of
each element at each value of the mass cut (defined as the 
mass that initially collapsed to form the compact remnant). 
For more details of the models, see \citet{tom07}.

The assumptions regarding the initial mass and the post-SN orbital 
distance of the secondary star are not so relevant due to the free 
parameter $f_{\rm cap}$. 
A different value for the initial mass of the secondary 
star, as small as (say) 1~\Msun with an initial radius of 
$R_{2,i} \approx 1.3$~\Rsuno at the time of the SN explosion, 
and for the post-SN orbital separation, say $a_{c,i} = 25$~\Rsuno, 
would require a larger capture efficiency factor of 
$f_{\rm cap} \approx 0.45$ (i.e., 45\%). 

The explosion energy is $E_K = 1 \times 10^{51}$ erg and $E_K = 
30 \times 10^{51}$ erg for the SN and HN
models, respectively. This energy is deposited instantaneously in 
the central region of the progenitor core to generate a strong 
shock wave. The subsequent propagation of the shock wave is 
followed through a hydrodynamic code 
\citep[][and references therein]{ume02}. As in \citet{gon08b},
our model computations assume different mass cuts and fallback
masses, and a mixing factor of 1 which assumes that all fallback
matter is well mixed with the ejecta. The amount of fallback, $M_{\rm
fall}$, is the difference between the final remnant mass, $M_{{\rm
BH},i}$, and the initial remnant mass of the explosion, $M_{\rm cut}$.
We should recall here the ejected mass, $\Delta M$, which is equal to
$M_{\rm He}-M_{{\rm BH},i}$, where $M_{\rm He}$ is the mass of the He
core.

We use SN/HN models to provide us with the yields of the 
explosion before radiative decay of element species. We then 
compute the integrated, decayed yields of the ejecta by adopting 
a mass cut and by mixing all of the material above the mass cut. 
Finally, we calculate the composition of the matter captured by
the secondary star, and we mix it with the material of its convective
envelope. It has been suggested that the black hole in 
the LMXB \mbox{Nova Sco 1994} could have formed in a two-stage 
process where the initial collapse led to the formation of a 
neutron star accompanied by a substantial kick and the final mass 
of the compact remnant was achieved by matter that fell
back after the initial collapse \citep{pod02}. 
However, the black hole mass in that system has been estimated 
at $\sim 5.4$~\Msun \citep{bee02} or $\sim 6.6$~\Msun 
\citep[][see Table~\ref{tgrav}]{sha03}, 
while the black hole in \mbox{V404 Cygni} may have an initial mass 
of $\sim 8$~\Msun which would require a fallback 
mass of $\sim 6.6$ \Msun if we assume $\sim 1.4$~\Msun for a canonical
neutron star. \citet{mac01} proposed a scenario where 
collapsar models harbor black holes that could form in a mild
explosion with substantial fallback (up to $\sim 5$~\Msun). 

\begin{figure*}[ht!]
\centering
\includegraphics[width=8.cm,angle=0]{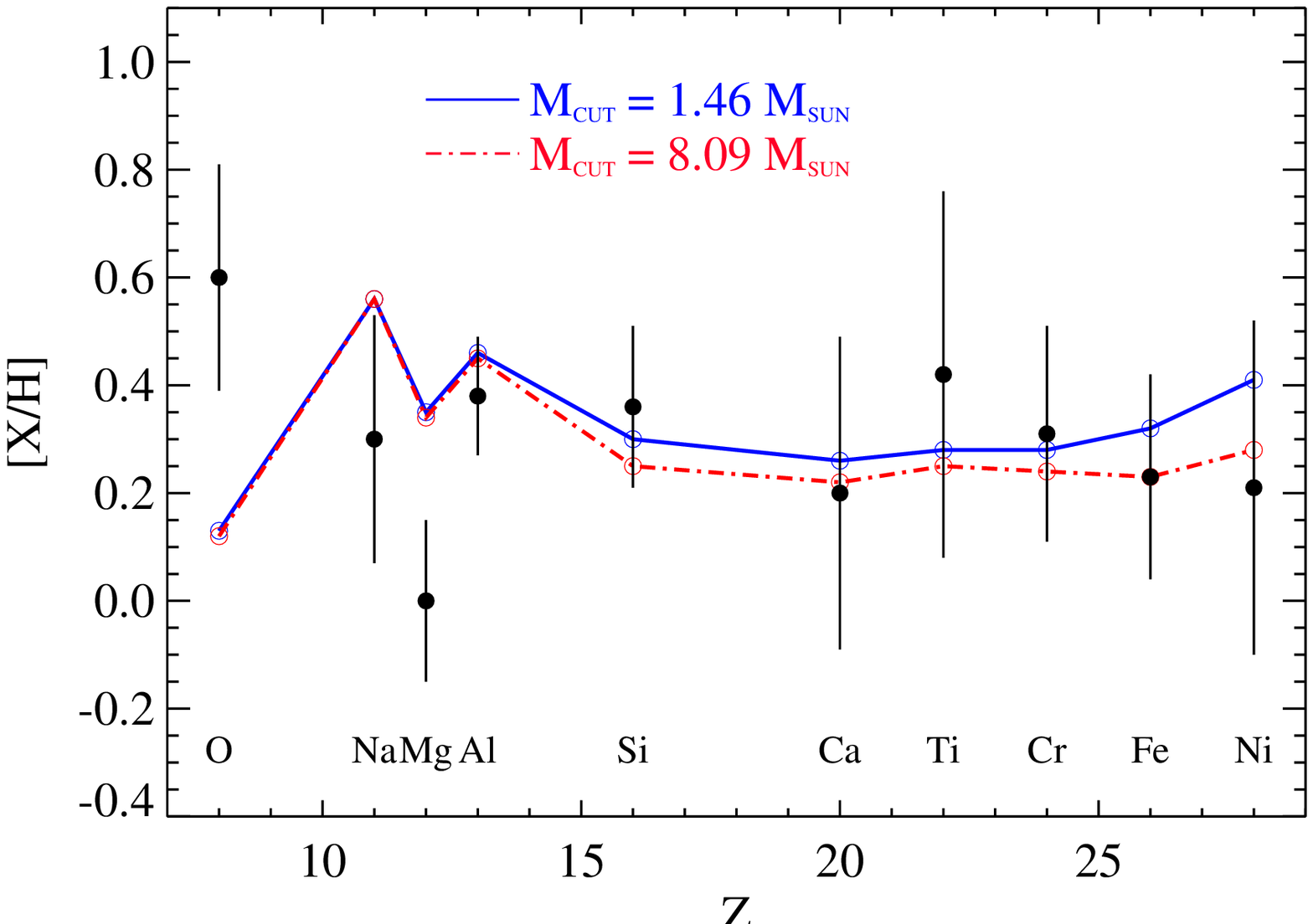}
\includegraphics[width=8.cm,angle=0]{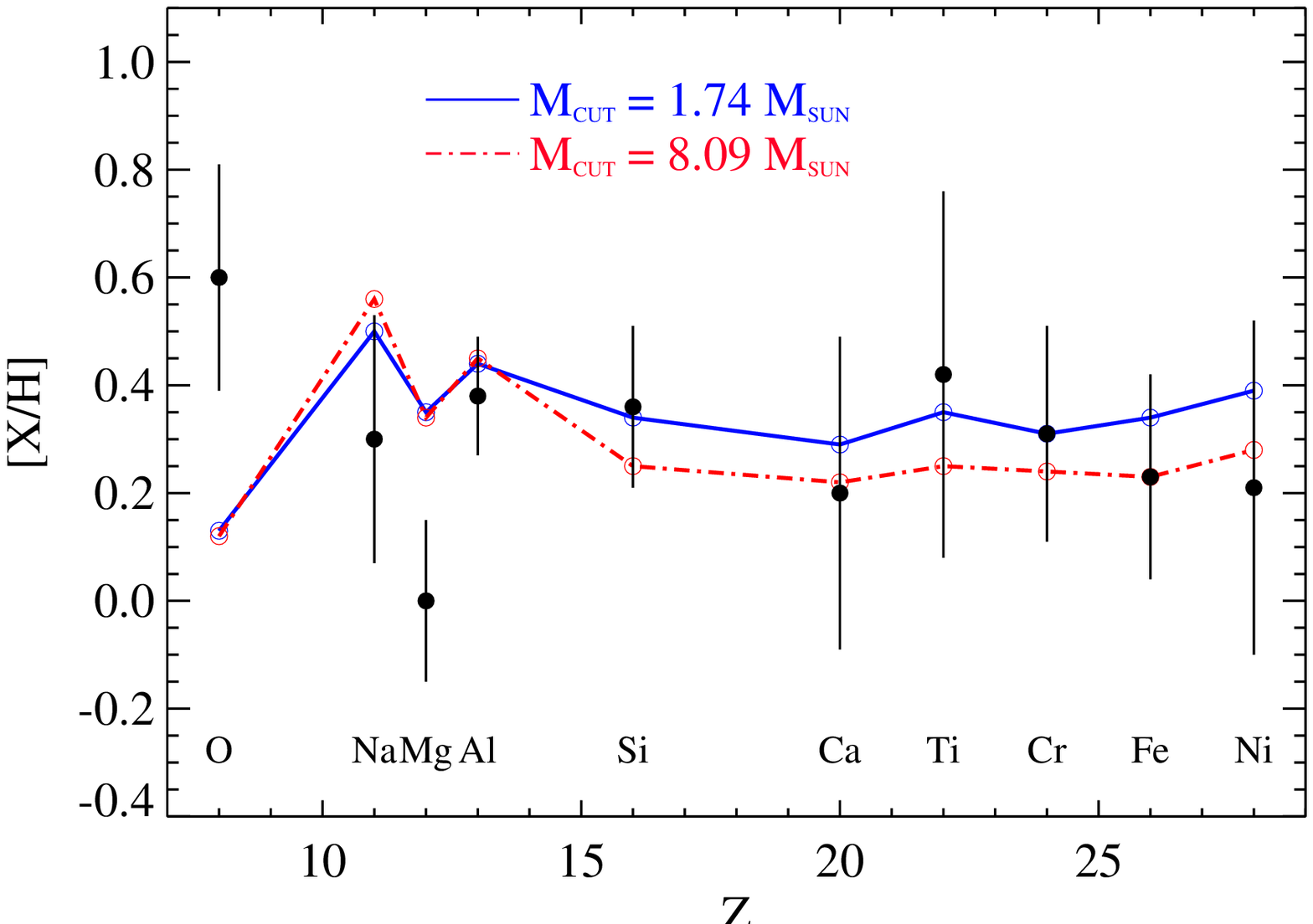}
\caption{\footnotesize{\emph{Left panel}: Observed abundances (filled circles
with error bars) in comparison with the expected abundances in the
secondary star of \mbox{V404 Cygni} after having captured
10\% of the matter ejected within the solid angle subtended 
by the secondary from a solar metallicity ($Z = 0.02$) 40~\Msun 
spherically symmetric supernova explosion 
($M_{\rm He} = 15.1$~\Msuno) with $E_K = 1 \times 10^{51}$ erg 
for two different mass cuts, $M_{\rm cut} = 1.46$~\Msun 
(solid line with open circles)
and $M_{\rm cut} = 8.09$~\Msun (dashed-dotted line with open circles).
The initial abundances of the secondary star were adopted for the
average abundances of thin-disk solar-type stars with 
[Fe/H]~$=0.23 \pm 0.12$.
\emph{Right panel}: Same as left panel, but for a spherically symmetric 
hypernova explosion ($E_K = 30 \times 10^{51}$ erg) for two different
mass cuts, $M_{\rm cut} = 1.74$~\Msun (solid line with open circles)
and $M_{\rm cut} = 8.09$~\Msun (dashed-dotted line with open circles).}}
\label{fsph}
\end{figure*}

In Figure~\ref{fsph} we show the expected abundances of the secondary 
star after contamination from the nucleosynthetic products of the SN
explosion ($E_K = 1 \times 10^{51}$ erg) of a $M_{\rm He} \approx 
15$~\Msun progenitor star. The initial abundances of the 
secondary star have been estimated from the average abundances 
of thin-disk solar-type stars with [Fe/H] $\approx 0.23$ \citep{gon10b},
which are provided in Table~\ref{tsn}.
Note that for a given model, 
the Al abundance in the secondary star hardly depends on 
the mass cut, since Al forms in the outer layers of the explosion. 
Thus, once the capture efficiency is fixed, the Al
abundance in Figure~\ref{fsph} is almost constant.

The expected abundances of O, Mg, and Al in the secondary star 
after contamination from the SN ejecta remain mostly independent 
of the adopted mass cut, whereas other elements like Si, Ca, Ti, Fe,
and Ni are quite sensitive to the mass cut of the model. This appears
to be more clear in the HN model, where the higher explosion energy
enhances the amounts of the $\alpha$-process elements 
Si, Ca, and Ti. The Fe-group elements like Cr, Fe, and Ni are also
sensitive to the mass cut, especially in the HN model.

For both the SN model (left panel in Fig.~\ref{fsph}) and the HN model 
(right panel in Fig.~\ref{fsph}), the predicted abundances
agree with the observed abundances in the 
secondary star relatively well,
except for those of O and Mg. The expected 
Mg abundance is too high in comparison with the observed value 
since the adopted initial Mg abundance is already higher than 
the observed abundance (see Table~\ref{tsn}). 
The case of O is just the contrary; the initial abundance is 
so low that it is not possible to reach the observed O abundance when
fitting other element abundances (see Fig.~\ref{fsph}).

For mass cuts above $\sim 3$~\Msuno, very little Si, Ca, Ti, Fe, Cr, 
and Ni is ejected; thus, the expected abundances of the model 
essentially reflect the initial abundances of the secondary star. 
In contrast, O, Na, Mg, and Al are hardly sensitive to the mass cut, 
and all are enhanced due to the capture of enriched material 
in the SN/HN explosion.

\begin{deluxetable}{lcccccccccccc}
\centering
\tabletypesize{\scriptsize}
\tablecolumns{7}
\tablecaption{Metal-Rich Supernova/Hypernova Explosion Models in
\mbox{V404 Cygni}}
\tablewidth{0pt}
\tablehead{ & & & \multicolumn{4}{c}{${\rm [X/H] \: Expected}$\tablenotemark{d}}\\
\cline{4-7}\\
Element & ${\rm [X/H]\:\rm Observed}$\tablenotemark{a} & ${\rm [X/H]}_{0}$\tablenotemark{b} &
$M_{\rm cut, low}$\tablenotemark{c} & $M_{\rm cut, up}$ & $M_{\rm cut, low}$ & $M_{\rm cut, up}$}
\startdata
 & & & \multicolumn{4}{c}{Spherical explosion model of $Z=0.02$} \\
\noalign{\smallskip}
\tableline
\noalign{\smallskip}
 & & & \multicolumn{2}{c}{Supernova} &
\multicolumn{2}{c}{Hypernova} \\
\noalign{\smallskip}
\cline{4-5} \cline{6-7} \\
\noalign{\smallskip}
 O & 0.60 & -0.09 &  0.13 &  0.12 &  0.13 &  0.12  \\
Na & 0.30 &  0.29 &  0.56 &  0.56 &  0.50 &  0.56  \\
Mg & 0.00 &  0.24 &  0.35 &  0.34 &  0.35 &  0.34  \\
Al & 0.38 &  0.27 &  0.46 &  0.45 &  0.44 &  0.45  \\
Si & 0.36 &  0.24 &  0.30 &  0.25 &  0.34 &  0.25  \\
Ca & 0.20 &  0.22 &  0.26 &  0.22 &  0.29 &  0.22  \\
Ti & 0.42 &  0.25 &  0.28 &  0.25 &  0.35 &  0.25  \\
Cr & 0.31 &  0.24 &  0.28 &  0.24 &  0.31 &  0.24  \\
Fe & 0.23 &  0.23 &  0.32 &  0.23 &  0.34 &  0.23  \\
Ni & 0.21 &  0.27 &  0.41 &  0.28 &  0.39 &  0.28  \\
\noalign{\smallskip}
\tableline
\noalign{\smallskip}
 & & & \multicolumn{4}{c}{Aspherical explosion model with $Z=0.02$}  \\
\noalign{\smallskip}
\tableline
\noalign{\smallskip}
 & & & \multicolumn{2}{c}{Angle\tablenotemark{e} $=0^\circ$--$15^\circ$}
 & \multicolumn{2}{c}{Angle $=0^\circ$--$90^\circ$} \\
\noalign{\smallskip}
\cline{4-5} \cline{6-7} \\
\noalign{\smallskip}
 O & 0.60 & -0.09 &  0.17 &  0.14 &  0.19 &  0.19  \\
Na & 0.30 &  0.29 &  0.39 &  0.38 &  0.38 &  0.39  \\
Mg & 0.00 &  0.24 &  0.36 &  0.35 &  0.36 &  0.37  \\
Al & 0.38 &  0.27 &  0.45 &  0.43 &  0.45 &  0.46  \\
Si & 0.36 &  0.24 &  0.31 &  0.28 &  0.39 &  0.33  \\
Ca & 0.20 &  0.22 &  0.24 &  0.22 &  0.33 &  0.27  \\
Ti & 0.42 &  0.25 &  0.25 &  0.25 &  0.41 &  0.35  \\
Cr & 0.31 &  0.24 &  0.24 &  0.24 &  0.33 &  0.29  \\
Fe & 0.23 &  0.23 &  0.23 &  0.23 &  0.31 &  0.26  \\
Ni & 0.21 &  0.27 &  0.27 &  0.27 &  0.40 &  0.33  \\
\enddata
\tablenotetext{a}{Observed abundances of the
secondary star in \mbox{V404 Cygni}.}

\tablenotetext{b}{Initial abundances assumed for
the secondary star in \mbox{V404 Cygni}; see text.}

\tablenotetext{c}{$M_{\rm cut, low}$ and $M_{\rm cut, up}$ are 
(respectively) the lower and upper mass cuts adopted in the model 
computation. See the exact value in the captions of Figs.~12 and 13.} 

\tablenotetext{d}{Expected abundances of the secondary star.}

\tablenotetext{e}{Angular range, measured from the equatorial plane,
in which all the ejected material in the explosion has been completely
mixed for each velocity point.} 

\tablecomments{Expected abundances in the secondary atmosphere
contaminated with nucleosynthetic products of metal-rich explosion
models for two different mass cuts and symmetries, presented
in Figs.~12 and 13.} 

\label{tsn}
\end{deluxetable}

We could have also investigated a model with solar initial abundances 
for the secondary star (${\rm [X/H]}_{0}=0$) and the same 
explosion model of solar metallicity ($Z=0.02$), although it is
unlikely that a significant amount of elements formed in the 
inner layers of the explosion (such as Ti, Ni, and Fe) would be 
present in the ejecta. In such models, only the HN explosion with 
a mass cut as low as $M_{\rm cut} \approx 2$~\Msun would fit the 
relatively enhanced observed abundances of Si, Ti, Cr, Fe, and Ni,
but we would still not be able to fit the O and Mg abundances. 
This low mass cut would require very efficient mixing processes. 
and the model would also need a small capture efficiency of 
$f_{\rm cap} \approx 0.15$ (i.e., only 15\% of the matter ejected 
within the solid angle subtended by the secondary star).  

\subsection{Aspherical Explosion}

\citet{mil09b} pointed out that due to the relatively small peculiar 
velocity of the system, \mbox{V404 Cygni} does not require an 
asymmetric kick. However, here we explore this 
possibility using explosion models 
from \citet{mae02} that are not spherically symmetric.
An aspherical SN explosion produces chemical 
inhomogeneities in the ejecta which are dependent on direction.
Thus, if the jet in the aspherical SN explosion is collimated 
perpendicular to the orbital plane of the binary 
\citep[for more details, see][]{gon05b}, where the 
secondary star is located, elements such as Ti, Fe, and Ni are 
ejected mainly in the jet direction, while O, Mg, Al, Si, and S 
are preferentially ejected near the equatorial plane of the 
helium star \citep{mae02}. 

\begin{figure*}
\centering
\includegraphics[width=8.cm,angle=0]{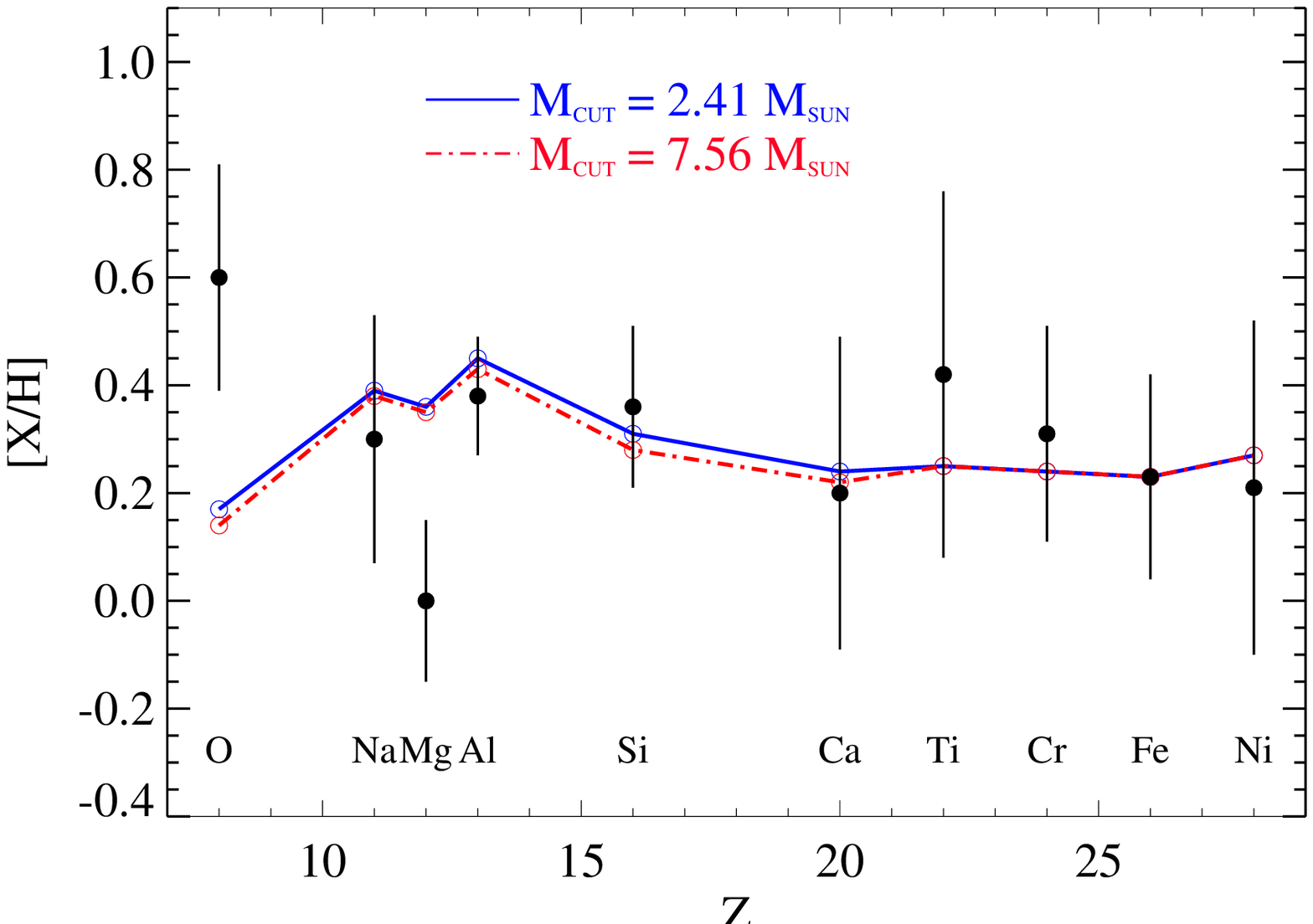}
\includegraphics[width=8.cm,angle=0]{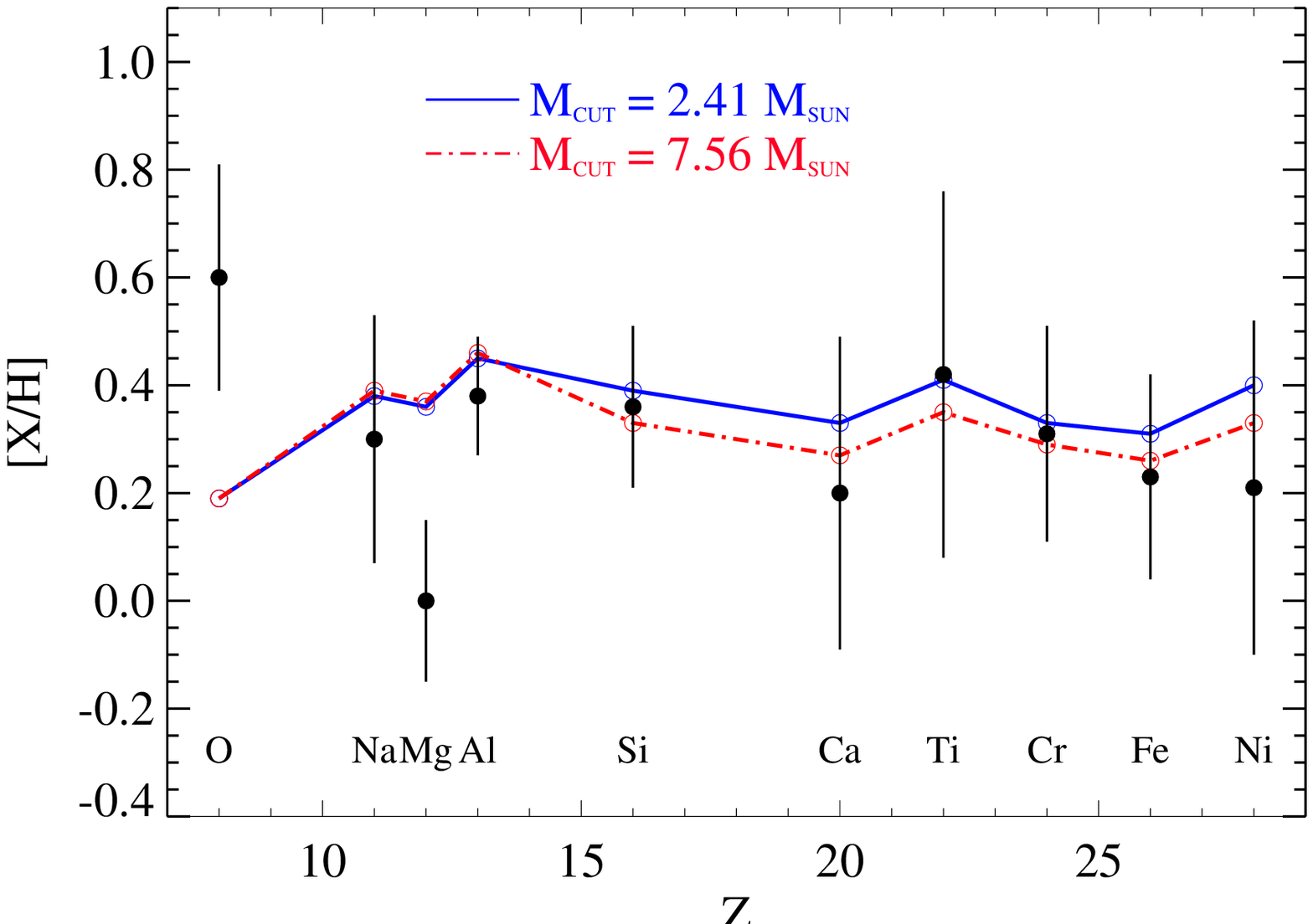}
\caption{\footnotesize{\emph{Left panel}: Observed abundances (filled circles
with error bars) in comparison with the expected abundances in the
secondary star  of \mbox{V404 Cygni} after having captured
$\sim 10$\% of the matter ejected within the solid angle subtended 
by the secondary from an aspherical SN explosion of $E_K = 10 \times 
10^{51}$ erg for two different mass cuts, $M_{\rm cut} = 2.41$~\Msun
(solid line with open circles) and $M_{\rm cut} = 7.56$~\Msun
(dashed-dotted line with open circles). This model corresponds to the
matter ejected in the equatorial plane of the primary, where we assumed
that the secondary star is located \citep[for more details, see][]{gon05b}.
\emph{Right panel}: Same as left panel, but in this model
we have assumed complete lateral mixing, where all of the material within
given velocity bins is completely mixed. Two simulations
are shown for two different mass cuts, $M_{\rm cut} = 2.41$~\Msun (solid
line with open circles) and $M_{\rm cut} = 7.56$~\Msun (dashed-dotted
line with open circles).}}  
\label{fasph}
\end{figure*}

In Figure~\ref{fasph} we compare the predicted abundances in the
atmosphere of the secondary star after pollution from an aspherical 
explosion model of a metal-rich progenitor having a 16~{\Msun}~He core 
(see also Table~\ref{tsn}). 
The initial abundances of the secondary were, as in the spherical case, 
equally extracted from the average values of solar-type stars of the solar 
neighborhood with similar iron content (see Table~\ref{tsn}). 
The left panel of Figure~\ref{fasph} reflects the composition of the 
material ejected in the equatorial plane, which should be rich in 
O, Na, Mg, Al, and Si --- the only elements 
significantly enhanced with respect to the initial abundances.
We see that the choice of the mass cut does not make a significant 
difference, and as in the spherical case, this model fits all of
the element abundances within their error bars except for O and Mg.

In the right panel of Figure~\ref{fasph} we have considered 
complete lateral mixing \citep{pod02} --- that is, the ejected matter 
is completely mixed within each velocity bin 
\citep[for more details, see][]{gon08b}. 
The observed abundances might be better reproduced if complete 
lateral mixing is adopted, since this process tends to enhance 
the Si, Ca, Ti, Cr, Fe, and Ni element abundances at all mass cuts.
A model with solar initial abundances for the
secondary star (i.e., ${\rm [X/H]}_{0}=0$) and the same explosion 
model of solar metallicity ($Z=0.02$) was also inspected, 
providing the same result except for the equatorial model, 
which produces too low abundances of Ti, Fe, and Ni in comparison 
with the observations.

\section{Conclusions}

We have presented Keck~I/HIRES high-resolution spectroscopy of the
black hole X-ray binary \mbox{V404 Cygni}. The spectra were obtained 
at system orbital phase close to zero to minimize the effect 
of the Roche-lobe symmetry of the secondary star. We have 
performed a detailed chemical analysis of the secondary star,
applying a technique that provides a determination of the stellar
parameters and takes into account any possible veiling from the
accretion disk. We find $T_{\mathrm{eff}} = 4800 \pm 100$~K, $\log 
[g/{\rm cm~s}^2] = 3.50 \pm 0.15$, $\mathrm{[Fe/H]} = 0.23 \pm 0.19$, 
and a disk veiling (defined as $F_{\rm  disk}/F_{\rm total}$) of
$\sim 5$\% at 5100~{\AA}, decreasing toward longer wavelengths.  

We have derived the chemical abundances of O, Na, Mg, Al,
Si, Ca, Ti, Cr, Fe, and Ni. They
are typically higher than solar, and some elements 
show additional slight enhancements (e.g., Al, Si, and Ti). 
The O abundance was derived using the \ion{O}{1} triplet 
at 7771-5~{\AA}, a robust O abundance indicator; it appears
to be a factor of 4 more abundant than in the Sun, and more than
a factor of 2 overabundant in comparison with
stars of the solar neighborhood having similar iron content.
These photospheric abundances suggest possible contamination from
nucleosynthetic products in the supernova/hypernova explosion that
formed the compact object in this system.

The peculiar velocity of the LMXB \mbox{V404 Cygni}, 
$v_{\rm pec}=39.9 \pm 5.5$~\kmso, is larger than the Galactic velocity 
dispersion, suggesting that the black hole in this system formed in a 
supernova event. 
Thus, the secondary star could have kept a record of 
this SN/HN event in the chemical composition of its atmosphere.
We explored this possibility using a variety 
of SN/HN explosion models for different geometries.
We compared the expected abundances in the 
companion star after contamination of nucleosynthetic products 
from initial metallicities of the secondary star adopted 
using the average values in solar-type thin-disk stars.
Metal-rich spherically symmetric models are able to reproduce 
the observed abundances relatively well, except for O and Mg, 
which appear to show
too high and too low (respectively) abundances in comparison to 
model predictions, regardless of the choice of the mass cut.

A spherical explosion easily provides the energy required to explain
the peculiar velocity of this system, which is in the Galactic
plane. Therefore, an asymmetric SN/HN explosion may
not be required. 
However, we have also modelled the expected abundances in the
secondary star after capturing a significant amount of material from
an aspherical explosion. These models also provide good
agreement for all elements, as in the spherical case, except for O 
and Mg, without invoking extensive fallback and mixing. 

In a forthcoming paper \citep[][in preparation]{gon11}, we will present
the Li abundance and study the $^6$Li/$^7$Li isotopic ratio to
explore different evolutionary scenarios and possible production of Li
due to the X-ray radiation in this black-hole binary system 
\citep[see][]{cas07b}.

\acknowledgments

We thank K. Maeda, H. Umeda, K. Nomoto, and N. Tominaga
for providing us with their spherical and aspherical explosion
models. 
We are grateful to T. Marsh for the use of the MOLLY analysis package.
J.I.G.H. acknowledges financial support from the Spanish Ministry of
Science and Innovation (MICINN) under the 2009 Juan de la Cierva
Programme.
J.I.G.H. and G.I. are grateful for financial
support from MICINN grant AYA2008-04874, while
J.C. acknowledges MICINN grant AYA2010-18080.
Additional funding was provided by MICINN grant 
AYA2005--05149, and (to A.V.F.) by US National Science Foundation grant
AST--0908886.
The W. M. Keck Observatory is operated as a scientific partnership 
among the California Institute of Technology, the University of 
California, and NASA; it was made possible by the generous financial 
support of the W. M. Keck Foundation. 
This work has made use of the VALD database and IRAF facilities.


\begin{thebibliography}{}

\bibitem[Allende Prieto et al.(2004)]{all04} 
Allende Prieto, C., Barklem, P.~S., Lambert, D.~L., \& 
Cunha, K.\ 2004, \aap, 420, 183 

\bibitem[Al-Naimiy(1978)]{aln78}
Al-Naimiy, H. M. 1978, \apss, 420, 183

\bibitem[Blaauw(1961)]{bla61} 
Blaauw, A.\ 1961, \bain, 15, 265 

\bibitem[Beer \& Podsiadlowski(2002)]{bee02}
Beer, M. E., \& Podsiadlowski, P. 2002, \mnras, 331, 351

\bibitem[Benz \& Hills(1992)]{bah92}
Benz, W., \& Hills, J. G. 1992, \apj, 433, 185

\bibitem[Brunish \& Truran(1982)]{bru82}
Brunish, W. M., \& Truran, J. W. 1982, \apjs, 49, 447 

\bibitem[Cantrell et al.(2010)]{can10} 
Cantrell, A.~G., et al.\ 2010, \apj, 710, 1127 

\bibitem[Casares et al.(1992)]{cas92} 
Casares, J., Charles, P.~A., \& Naylor, T.\ 1992, \nat, 355, 614 

\bibitem[Casares et al.(1993)]{cas93} 
Casares, J., Charles, P.~A., Naylor, T., \& Pavlenko, E.~P.\ 1993, 
\mnras, 265, 834 

\bibitem[Casares \& Charles(1994)]{cas94} 
Casares, J., \& Charles, P.~A.\ 1994, \mnras, 271, L5 

\bibitem[Casares(1996)]{cas96} 
Casares, J.\ 1996, IAU Colloq.~158: Cataclysmic Variables and 
Related Objects, 208, 395 

\bibitem[Casares(2007)]{cas07a} 
Casares, J.\ 2007, IAU Symposium, 238, 3 

\bibitem[Casares et al.(2007)]{cas07b} 
Casares, J., Bonifacio, P., Gonz{\'a}lez Hern{\'a}ndez, J.~I., 
Molaro, P., \& Zoccali, M.\ 2007, \aap, 470, 1033 

\bibitem[Chen(1997)]{che97} 
Chen, B. 1997, \apj, 491, 181 

\bibitem[Clayton(1983)]{cla83} 
Clayton, D.~D.\ 1983, Principles of Stellar Evolution and
Nucleosynthesis (Chicago: Univ. Chicago Press)

\bibitem[Ecuvillon et al.(2006)]{ecu06}
Ecuvillon, A., Israelian, G., Santos, N. C., Shchukina, N. G., Mayor,
M., \& Rebolo, R. 2006, \aap, 445, 633

\bibitem[Eggleton(1983)]{egg83} 
Eggleton, P.~P.\ 1983, \apj, 268, 368 

\bibitem[Froning et al.(2007)]{fro07} 
Froning, C.~S., Robinson, E.~L., \& Bitner, M.~A.\ 2007, \apj, 663, 1215 

\bibitem[Gelino et al.(2001)]{gel01}
Gelino, D. M., Harrison, T. E., \& Orosz, J. A. 2001, \apj, 122, 2668

\bibitem[Gelino et al.(2006)]{gel06}
Gelino, D. M., Balman, \c{S}., Kililo\u{g}lu, \"U., Yilmaz, A.,
Kalemci, E., \& Tomsick, J. A. 2006, \apj, 642, 438

\bibitem[Gonz\'alez Hern\'andez et al.(2004)]{gon04}
Gonz\'alez Hern\'andez, J. I., Rebolo, R., Israelian, G., Casares, J.,
Maeder, A., \& Meynet, G. 2004, \apj, 609, 988

\bibitem[Gonz{\'a}lez Hern{\'a}ndez et al.(2005)]{gon05a} 
Gonz{\'a}lez Hern{\'a}ndez, J.~I., Rebolo, R., Pe{\~n}arrubia, J., 
Casares, J., \& Israelian, G.\ 2005, \aap, 435, 1185  

\bibitem[Gonz\'alez Hern\'andez et al.(2005)]{gon05b}
Gonz\'alez Hern\'andez, J. I., Rebolo, R., Israelian, G., Casares, J.,
Maeda, K., Bonifacio, P., \& Molaro, P. 2005, \apj, 630, 495

\bibitem[Gonz\'alez Hern\'andez et al.(2006)]{gon06}
Gonz\'alez Hern\'andez, J. I., Rebolo, R., Israelian, G., Harlaftis, E.
T., Filippenko, A. V., \& Chornock, R. 2006, \apj, 644, L49

\bibitem[Gonz{\'a}lez Hern{\'a}ndez et al.(2007)]{gon07} 
Gonz{\'a}lez Hern{\'a}ndez, J.~I., Rebolo, R., 
\& Israelian, G.\ 2007, IAU Symposium, 238, 43 

\bibitem[Gonz{\'a}lez Hern{\'a}ndez et al.(2008a)]{gon08a} 
Gonz{\'a}lez Hern{\'a}ndez, J.~I., Rebolo, R., \& 
Israelian, G.\ 2008a, \aap, 478, 203 

\bibitem[Gonz{\'a}lez Hern{\'a}ndez et al.(2008b)]{gon08b} 
Gonz{\'a}lez Hern{\'a}ndez, J.~I., Rebolo, R., Israelian, G., 
Filippenko, A.~V., Chornock, R., Tominaga, N., Umeda, H., 
\& Nomoto, K.\ 2008b, \apj, 679, 732 

\bibitem[Gonz{\'a}lez Hern{\'a}ndez et al.(2009)]{gon09} 
Gonz{\'a}lez Hern{\'a}ndez, J.~I., Iglesias-Groth, S., Rebolo, R., 
Garc{\'{\i}}a-Hern{\'a}ndez, D.~A., Manchado, A., 
\& Lambert, D.~L.\ 2009, \apj, 706, 866 

\bibitem[Gonz{\'a}lez Hern{\'a}ndez \& Casares(2010)]{gon10a} 
Gonz{\'a}lez Hern{\'a}ndez, J.~I., \& Casares, J.\ 2010, \aap, 516, A58 

\bibitem[Gonz{\'a}lez Hern{\'a}ndez et al.(2010)]{gon10b} 
Gonz{\'a}lez Hern{\'a}ndez, J.~I., Israelian, G., Santos, N.~C., 
Sousa, S., Delgado-Mena, E., Neves, V., \& Udry, S.\ 2010, \apj, 720, 1592 

\bibitem[Gonz{\'a}lez Hern{\'a}ndez et al.(2011)]{gon11} 
Gonz{\'a}lez Hern{\'a}ndez, J.~I., Casares, J., Rebolo, R., 
Israelian, G., Filippenko, A.~V., \& Chornock, R. 2011, \apj, in
preparation

\bibitem[Greene et al.(2001)]{gree01} 
Greene, J., Bailyn, C.~D., \& Orosz, J.~A.\ 2001, \apj, 554, 1290 

\bibitem[Grevesse et al.(1996)]{gre96}
Grevesse, N., Noels, A., \& Sauval, A. J. 1996, in Cosmic Abundances, 
ed. S. S. Holt \& G. Sonneborn (San Francisco: ASP, Conf. Ser. Vol. 
99), 117

\bibitem[Haswell et al.(2002)]{has02}
Haswell, C. A., Hynes, R. I., King, A. R., \& Schenker, K. 2002,
\mnras, 332, 928

\bibitem[Hills(1983)]{hil83}
Hills, J. G. 1983, \apj, 267, 322

\bibitem[Hills(1991)]{hil91}
Hills, J. G. 1991, \apj, 102, 2

\bibitem[Hjellming \& Rupen(1995)]{hje95} 
Hjellming, R.~M., \& Rupen, M.~P.\ 1995, \nat, 375, 464 

\bibitem[Hynes et al.(2005)]{hyn05}
Hynes, R. I., Robinson, E. L., \& Bitner, M. 2005, \apj, 630, 405

\bibitem[Hynes et al.(2009)]{hyn09} 
Hynes, R.~I., Bradley, C.~K., Rupen, M., Gallo, E., Fender, R.~P., 
Casares, J., \& Zurita, C.\ 2009, \mnras, 399, 2239 

\bibitem[Israelian et al.(1999)]{isr99}
Israelian, G., Rebolo, R., Basri, G., Casares, J., \&
Mart{\'\i}n, E. L. 1999, \nat, 401, 142

\bibitem[Jonker \& Nelemans(2004)]{jon04} 
Jonker, P.~G., \& Nelemans, G.\ 2004, \mnras, 354, 355 

\bibitem[Khargharia et al.(2010)]{kha10} 
Khargharia, J., Froning, C.~S., \& Robinson, E.~L.\ 2010, \apj, 716, 
1105 

\bibitem[Kifonidis et al.(2000)]{kif00}
Kifonidis, K., Plewa, T., Janka, H.-Th., \& M\"{u}ller, E. 2000, 
\aap, 531, L123 

\bibitem[Kurucz et al.(1993)]{kur93}
Kurucz, R. L. ATLAS9 Stellar Atmospheres Programs and 2 \kms
Grid. (CD-ROM, Smithsonian Astrophysical Observatory, Cambridge,
1993).

\bibitem[Kurucz et al.(1984)]{kur84}
Kurucz, R. L., Furenild, I., Brault, J., \& Testerman, L. 1984, Solar 
Flux Atlas from 296 to 1300 nm, NOAO Atlas 1 (Cambridge: Harvard Univ. 
Press)

\bibitem[Lai et al.(2001)]{lai01}
Lai, D., Chernoff, D. F., \& Cordes, J. M. 2001, \apj, 549, 1111

\bibitem[MacFadyen et al.(2001)]{mac01}
MacFadyen, A. I., Woosley, S. E., \& Heger, A. 2001, \apj, 550, 410

\bibitem[Maeda et al.(2002)]{mae02}
Maeda, K., Nakamura, T., Nomoto, K., Mazzali, P. A., Patat, F., \&
Hachisu, I. 2002, \apj, 565, 405

\bibitem[Marsh et al.(1994)]{mah94}
Marsh, T. R., Robinson, E. L., \& Wood, J. H. 1994, \mnras, 266, 137
 
\bibitem[Mignard(2000)]{mig00} 
Mignard, F.\ 2000, \aap, 354, 522 

\bibitem[Miller-Jones et al.(2009a)]{mil09a} 
Miller-Jones, J.~C.~A., Jonker, P.~G., Nelemans, G., Portegies Zwart, 
S., Dhawan, V., Brisken, W., Gallo, E., \& Rupen, M.~P.\ 2009a, 
\mnras, 394, 1440 

\bibitem[Miller-Jones et al.(2009b)]{mil09b} 
Miller-Jones, J.~C.~A., Jonker, P.~G., Dhawan, V., Brisken, W.,
Rupen, M.~P., Nelemans, G., \& Gallo, E.\ 2009b, \apjl, 706, L230 

\bibitem[Mirabel et al.(2001)]{mir01} 
Mirabel, I.~F., Dhawan, 
V., Mignani, R.~P., Rodrigues, I., 
\& Guglielmetti, F.\ 2001, \nat, 413, 139 

\bibitem[Mirabel et al.(2002)]{mir02} 
Mirabel, I.~F., Mignani, R., Rodrigues, I., Combi, J.~A., 
Rodr{\'{\i}}guez, L.~F., \& Guglielmetti, F.\ 2002, \aap, 395, 595 

\bibitem[Neilsen et al.(2008)]{nei08} 
Neilsen, J., Steeghs, D., \& Vrtilek, S.~D.\ 2008, \mnras, 384, 849 

\bibitem[Orosz et al.(2001)]{oro01}
Orosz, J. A., et al. 2001, \apj, 555, 489 

\bibitem[Orosz \& Bailyn(1997)]{oro97} 
Orosz, J.~A., \& Bailyn, C.~D.\ 1997, \apj, 477, 876 

\bibitem[Pavlenko et al.(1996)]{pav96} 
Pavlenko, E.~P., Martin, A.~C., Casares, J., Charles, P.~A., 
\& Ketsaris, N.~A.\ 1996, \mnras, 281, 1094 

\bibitem[Piskunov et al.(1995)]{pis95}
Piskunov, N. E., Kupka, F., Ryabchikova, T. A., Weiss, W. W., \& Jeffery, C.
S. 1995, \aaps, 112, 525

\bibitem[Podsiadlowski et al.(2002)]{pod02}
Podsiadlowski, P., Nomoto, K., Maeda, K., Nakamura, T., Mazzali, P., \&
Schmidt, B. 2002, \apj, 567, 491 

\bibitem[Portegies Zwart et al.(1997a)]{por97a} 
Portegies Zwart, S.~F., Verbunt, F., \& Ergma, E.\ 1997a, \aap, 321, 207 

\bibitem[Portegies Zwart et al.(1997b)]{1997b} 
Portegies Zwart, S.~F., Kouwenhoven, M.~L.~A., \& Reynolds, A.~P.\
1997b, \aap, 328, L33  

\bibitem[Sanwal et al.(1996)]{san96} 
Sanwal, D., Robinson, E.~L., Zhang, E., Colome, C., Harvey, P.~M., 
Ramseyer, T.~F., Hellier, C., \& Wood, J.~H.\ 1996, \apj, 460, 437 

\bibitem[Sadakane et al.(2006)]{sad06}
Sadakane, K., et al. 2006, \pasj, 58, 595 

\bibitem[Shahbaz et al.(1994)]{sha94} 
Shahbaz, T., Ringwald, F.~A., Bunn, J.~C., Naylor, T., 
Charles, P.~A., \& Casares, J.\ 1994, \mnras, 271, L10 

\bibitem[Shahbaz et al.(1999)]{sha99} 
Shahbaz, T., van der Hooft, F., Casares, J., Charles, P.~A., 
\& van Paradijs, J.\ 1999, \mnras, 306, 89 

\bibitem[Shahbaz(2003)]{sha03} 
Shahbaz, T.\ 2003, \mnras, 339, 1031 

\bibitem[Sneden(1973)]{sne73}
Sneden, C. 1973,  PhD Dissertation, Univ. of Texas at Austin

\bibitem[Tominaga et al.(2007)]{tom07} 
Tominaga, N., Umeda, H. \& Nomoto, K. 2007, \apj, 660, 516 

\bibitem[Torres et al.(2002)]{tor02} 
Torres, M.~A.~P., Casares, J., Mart{\'{\i}}nez-Pais, I.~G., 
\& Charles, P.~A.\ 2002, \mnras, 334, 233 

\bibitem[Torres et al.(2004)]{tor04} 
Torres, M.~A.~P., Callanan, P.~J., Garcia, M.~R., Zhao, P., Laycock, S., 
\& Kong, A.~K.~H.\ 2004, \apj, 612, 1026 

\bibitem[Umeda \& Nomoto(2002)]{ume02} 
Umeda, H., \& Nomoto, K.\ 2002, \apj, 565, 385 

\bibitem[Umeda \& Nomoto(2005)]{ume05} 
Umeda, H., \& Nomoto, K.\ 2005, \apj, 619, 427 

\bibitem[van Belle et al.(1999)]{bel99} 
van Belle, G.~T., et al.\ 1999, \aj, 117, 521 

\bibitem[van den Heuvel \& Habets(1984)]{heu84} 
van den Heuvel, E.~P.~J., \& Habets, G.~M.~H.~J.\ 1984, \nat, 309, 598 

\bibitem[van der Hooft et al.(1998)]{hoo98} 
van der Hooft, F., Heemskerk, M.~H.~M., Alberts, F., \& 
van Paradijs, J.\ 1998, \aap, 329, 538 

\bibitem[Vogt et al.(1994)]{vog94} 
Vogt, S.~S., et al.\ 1994, \procspie, 2198, 362 

\bibitem[Zhao et al.(1998)]{zha98} 
Zhao, G., Butler, K., \& Gehren, T.\ 1998, \aap, 333, 219 

\bibitem[Zhao \& Gehren(2000)]{zha00} 
Zhao, G., \& Gehren, T.\ 2000, \aap, 362, 1077 


\end{thebibliography}
\end{document}